\documentclass[times,sort&compress,3p]{elsarticle}
\usepackage[labelfont=bf]{caption}

\usepackage{amsmath,amsfonts,amssymb,amsthm,booktabs,color,epsfig,graphicx,hyperref,url}

\usepackage{amsthm}
\usepackage{verbatim,color,amssymb}
\usepackage{amsmath}					
\usepackage{amsthm}					
\usepackage{setspace}
\usepackage[mathscr]{euscript}
\usepackage{fancyhdr}
\usepackage{enumitem}
\usepackage{graphicx}
\usepackage{geometry}
\usepackage{lineno}
\usepackage{listings}
\usepackage{rotating}
\usepackage{subfig,subfloat}
\usepackage{booktabs}

\usepackage{mathtools}
\usepackage{subfig}
\usepackage{caption}
\usepackage{xr}
\usepackage{xurl}
\usepackage{float}
\newsubfloat{figure}
\usepackage{tikz}
\usetikzlibrary{arrows,chains,backgrounds,fit}
\usepackage{multirow}
\usepackage{lineno}

\theoremstyle{plain}
\newtheorem{theorem}{Theorem}

\theoremstyle{definition}

\def\bzero{{\mathbf 0}}
\newcommand{\uzero}            {\mbox{\boldmath$0$}}

\def\diag{\hbox{diag}}

\def\diag{\hbox{diag}}
\def\log{\hbox{log}}

\def\var{\hbox{var}}

\def\trace{\hbox{trace}}

\def\Beta{\hbox{Beta}}

\def\MVN{\hbox{MVN}}

\def\diag{\hbox{diag}}

\def\diag{\hbox{diag}}
\def\log{\hbox{log}}

\def\refhg{\hangindent=20pt\hangafter=1}
\def\refmark{\par\vskip 2mm\noindent\refhg}

\def\var{\hbox{var}}

\def\trace{\hbox{trace}}
\def\refhg{\hangindent=20pt\hangafter=1}
\def\refmark{\par\vskip 2mm\noindent\refhg}

\def\Beta{\hbox{Beta}}

\def\P_25_ICML{{\it Proceedings of the 25th international conference on Machine learning}}

\def\refhg{\hangindent=20pt\hangafter=1}
\def\refmark{\par\vskip 2mm\noindent\refhg}

\def\refhg{\hangindent=20pt\hangafter=1}
\def\refmark{\par\vskip 2mm\noindent\refhg}

\def\bse{\begin{eqnarray*}}
	\def\ese{\end{eqnarray*}}
\def\be{\begin{eqnarray}}
\def\ee{\end{eqnarray}}
\def\bq{\begin{equation}}
\def\eq{\end{equation}}

\def\boldpsi{{\mbox{\boldmath $\psi$}}}

\def\trans{^{\rm T}}

\def\bone{{\mathbf 1}}

\def\b1e{{\mathbf e}}

\def\bq{{\mathbf q}}

\def\bzero{{\mathbf 0}}

\newcommand{\bSigma}{\mbox{\boldmath $\Sigma$}}

\newcommand{\uB}       {\mbox{\boldmath$B$}}

\newcommand{\uD}       {\mbox{\boldmath$D$}}

\newcommand{\uF}       {\mbox{\boldmath$F$}}

\newcommand{\uI}       {\mbox{\boldmath$I$}}

\newcommand{\uJ}       {\mbox{\boldmath$J$}}

\newcommand{\uN}       {\mbox{\boldmath$N$}}

\newcommand{\uS}       {\mbox{\boldmath$S$}}

\newcommand{\uX}       {\mbox{\boldmath$X$}}

\newcommand{\uY}       {\mbox{\boldmath$Y$}}

\newcommand{\ugamma}            {\mbox{\boldmath$\gamma$}}
\newcommand{\udelta}            {\mbox{\boldmath$\delta$}}
\newcommand{\uepsilon}          {\mbox{\boldmath$\epsilon$}}

\newcommand{\uiota}             {\mbox{\boldmath$\uiota$}}

\newcommand{\umu}               {\mbox{\boldmath$\mu$}}

\newcommand{\uSigma}            {\mbox{\boldmath$\Sigma$}}

\newcommand{\uPhi}              {\mbox{\boldmath$\Phi$}}

\begin{document}

\begin{frontmatter}

\title{An approximate Bayes factor based high dimensional MANOVA using random projections}

\author[1]{Roger S. Zoh \corref{mycorrespondingauthor}}
\author[2]{Fangzheng Xie}

\address[1]{Department of Epidemiology \& Biostatistics, Indiana University, Bloomington, IN 47405, USA}
\address[2]{Department of Statistics, Indiana University, Bloomington, IN 47408, USA}

\cortext[mycorrespondingauthor]{Corresponding author. Email address: \url{rszoh@iu.edu}}

\begin{abstract}
High-dimensional mean vector testing problems for two or more independent groups remain a very active research area. When the length of the vector mean exceeds the groups' combined sample sizes, traditional tests are not applicable since they involve the inversion of rank deficient sample covariance matrices. Most approaches considered in the literature overcome this limitation by imposing a structure on the covariance matrices. Unfortunately, these assumptions are often unrealistic and difficult to justify in practice. 
We develop a Bayes factor (BF)-based testing procedure for comparing two or more population means in (very) high dimensional settings while making no a priori assumptions about the structure of the large unknown covariance matrices. Our test is based on random projections (RPs), a popular data perturbation technique. RPs are appealing since they make no assumptions about the form of the dependency across features in the data. Two versions of the Bayes factor-based test statistics are considered. As is common with data perturbation techniques, tests based on a single random projection can be misleading. Thus, our final test statistic is based on an ensemble of Bayes factors corresponding to multiple replications of randomly projected data. Both proposed test statistics are compared through a battery of simulation settings. Finally they are applied to the analysis of a publicly available single cell RNA-seq (scRNA-seq) dataset. 
\end{abstract}


\end{frontmatter}

\section{Introduction} \label{sec:intro}
The problem of comparing multiple group means continues to receive considerable attention in the literature, especially in the ‘large-p-small-n’ setting where $p >> n$. For the multiple sample testing problem, the approaches proposed in literature all center on a version of the Hotelling's $T^2$ statistic. Namely, the statistic used is
\be
T^{2} = C_{n}(\overline{\uX} - \overline{\uY})\trans\uS^{-1} (\overline{\uX} - \overline{\uY}), \label{eq:eqT2}
\ee
where $C_{n}$ is free a data-free quantity, $\uS$ is the (pooled) sample covariance, and the sample mean vectors are $\overline{\uX}$ and $\overline{\uY}$.
Unfortunately, in its original form (\ref{eq:eqT2}), the $T^2$ statistic can quickly become ill formed since it involves the inversion of a sample covariance matrix that is not positive definite when the dimension of the vector exceeds the combined sample size. Various approaches exist in literature to help circumvent these limitations. Solutions to that problem have centered around the following approaches. One approach ignores dependency between the features or groups of features \cite{ahmad2014u,bai1996effect,chen2010two,feng2017composite}. This has the direct effect of removing the issue of inverting ill-formed covariance matrices. The second approach can be viewed as a regularization scheme with a goal of making the sample covariance invertible. Two regularization schemes have emerged \cite{hu2020pairwise}. One regularization scheme uses a ridge type estimator for the sample covariance matrix \cite{chen2011regularized,li2020adaptable}. Another regularization approach, which is in principle closer to a data perturbation approach than a regularization approach as we commonly know it, is based on a random projection. Basically, random projection approach works by projecting the originally high dimensional data to a low-dimensional embedding and performing the test with these lower dimension data, which completely eliminates the need to invert a rank degenerate sample covariance matrix. This approach includes versions for both frequentist \cite{lopes2011more,srivastava2014raptt,thulin2014high}) and Bayesian \cite{zoh2018powerful} settings. Recently, there has been a growing effort towards combining these two approaches in the two-group mean testing problem \cite{hu2020pairwise}.

The two-sample mean testing problem in high dimensional settings is a special case of the more general MANOVA (Multivariate Analysis of Variance) problem. However, extending two-group mean testing procedures to testing more than two groups is not a trivial task \cite{cai2014high}. Suppose $G$ populations of dimension $p$, with the mean vector specified respectively as $\umu_1, \cdots, \umu_{G}$, and the common covariance matrix as $\uSigma$. In MANOVA, the testing problem is formulated as
\be
H_{0}:\; \umu_i = \umu_j\; \; \forall \; (i,j) \in \mathcal{P}\;  \; \mbox{versus} \; \; H_{1}:\; \exists \; (i,j) \in \mathcal{P}\; \mbox{s.t}   \; \umu_i \neq \umu_j\;  \label{eq:test1}
\ee
where $\mathcal{P} = \{(i,j): 1 \leq i < j \leq G  \}$.
Work on the more general (more than two groups) MANOVA approach when $p$ exceeds the sample sizes began more than 60 years ago \cite{dempster1958high,dempster1960significance}. In general, the approaches in the literature rely on one of two major assumptions. One approach derives the test under the assumption of common variances across groups \cite{fujikoshi2004asymptotic}. Another approach removes the assumption of common covariances \cite{srivastava2007multivariate}.

To our knowledge, random projections have not yet been considered for multiple group mean testing in high dimensions. Recently, random matrix approaches in general and random projections (RP) in particular have emerged as effective (linear) data reduction techniques in many fields \cite{wan2020sharp,lopez2021tuning}. Additionally, RPs have already proven very successful for two-group mean tests \cite{lopes2011more, srivastava2014raptt,zoh2018powerful}. However, to the best of our knowledge, RPs have not been used or evaluated in MANOVA for testing means of more than two groups. The goal of this paper is to investigate the performance of RPs in a Bayes factor-based test for MANOVA. The paper is structured as follows. In Section~\ref{sec:test}, we derive the Bayes factor-based tests. Section~\ref{sec:theori} provides theoretical results of our test along with simulation results. In Section~\ref{sec:Application}, we apply the proposed method to the analysis of an actual data set from single cell sequencing (scRNA-seq). We end with some concluding remarks in Section~\ref{sec:conclusion}. 
 

\section{Bayes factor-based tests} \label{sec:test}
Suppose the following data generating model: $\uX_{ig} = \umu_i + \uepsilon_{ig}$, $g = 1, \cdots, G$, with $G$ denoting the number of independent groups under consideration.
Note here that we assume that $G \geq 2$ and $\uepsilon_{ig} \stackrel{iid}{\sim} \MVN_{p}(\bzero, \uSigma)$; $\MVN_{p}$ denotes a multivariate-normal distribution with dimension $p$, mean vector $\bzero$ and positive-definite covariance matrix $\uSigma$. Suppose the following data matrices are observed (independently) for each of the $G$ groups as $\uX_1 \in \mathbb{R}^{n_1 \times p}, \cdots,  \uX_G \in \mathbb{R}^{n_G \times p}$, where the data vectors are stacked row-wise for all $n_g$ individuals in group $g$. 
Let $\udelta_{ij} = \umu_i - \umu_j$. The compound hypothesis in \eqref{eq:test1} can thus be expressed as
\be
H_0:\; \udelta_{ij} &=& \uzero\;\;\forall (i,j) \in \mathcal{P} \nonumber \\
 &\mbox{vs.}& \nonumber \\
 H_1: \; \udelta_{ij}  &\neq & \uzero\;\;\mbox{for at least one pair}\;\;(i,j) \in \mathcal{P},\;\; \label{eq:hyp1} 
\ee
which is equivalent to performing $|\mathcal{P}| = G(G-1)/2$ (cardinality of $\mathcal{P}$) pairwise comparisons and similar to prior approaches \cite{ahmad2014u,tony2014two}. To obtain the Bayes factor, we specify the prior for $\udelta_{ij}$ under the alternative ($H_1$) as $\udelta_{ij} \sim \MVN_{p}(\uzero, \uSigma/\tau_{ij})$, where $\uSigma$ is the covariance matrix common to all groups and $ 0 < \tau_{ij} < \infty$ is a positive constant scaling factor. Finally, since the common covariance matrix $\uSigma$ is unknown, both under the null, $H_0$, and the alternative, $H_1$, computing the Bayes factor requires a prior for the covariance matrix $\uSigma$. Various distributions for positive definite covariance matrices can be considered such as the Inverse-Wishart or the Matrix-F \cite{mulder2018matrix}. The choice of prior is often balanced between computational tractability and strength of the assumed prior on the analysis. To that end, we choose an noninformative prior for the covariance matrix by assuming a Jeffrey's prior for the covariance matrix with density proportional to $P(\uSigma) \propto |\uSigma|^{-(p+1)/2}$. 
Although MANOVA tests proposed in high-dimensional settings for the most part bypass the inversion of ill-formed sample covariance matrices, we choose instead to transform the high-dimensional testing problem into a lower-dimensional one while preserving (or minimally disturbing) the dependencies between the vector coordinates. Thus, our test uses a Bayes factor centered on the commonly used Hotelling $T^{2}$ statistic. We discuss in detail the two variants of the Bayes factor tests we considered.

\subsection{Bayes factor-based on the pooled covariance matrix ($BF^{PL}_{}$)} \label{sec:testpl}
For the case when $G = 2$ in high-dimensional settings with $p >> n_1+n_2$, in  \cite{zoh2018powerful}, we proposed a test based on a Bayes factor (in favor of the alternative) using random projections (RPs). This test, involving an RP matrix $\uPhi \in \mathbb{R}^{p \times m}$, is defined as
\be
BF_{12}(\uPhi) &=& \left(1 + \eta_{12} \right)^{-m/2} \left\{\frac{  1 + \frac{m f_{1,2}}{(1 + \eta_{12})(n-m-G+1)}}{ 1 + \frac{m f_{1,2}}{(n-m-G+1)}  } \right\}^{-(n-1)/2}, \label{eq:BF1}
\ee
where $\eta_{12} = n_{0,12}/\tau$, $f_{1,2}  = \frac{n-m-G+1}{(n-G)m} n_{0,12} (\overline{\uX}_1 - \overline{\uX}_2)\trans\uPhi(\uPhi\trans\uS_{p}\uPhi\trans)^{-1}\uPhi\trans(\overline{\uX}_1 - \overline{\uX}_2)$, $n = n_1 + n_2$,
 $ 1/n_{0,12} = 1/n_1 + 1/n_2$, $\overline{\uX}_g$ is group $g$ sample mean, $\uS_p = \sum_{g=1}^{2} (n_g -1)\uS_{p, g}/(n_1 +n_2-2)$ is the $p \times p$ pooled sample covariance matrix, $\uS_{p,g}$ is the sample covariance for group $g$, and $m$ is the lower dimensional projection space chosen so that $m < n_1 +n_2 - 2 << p$. We have described an approach to selecting $m$ and $\tau$ in \cite{zoh2018powerful}. Note here that $m$ depends on $n_1$ and $n_2$ but is totally independent of $p$.
 Similarly, to test the (complex) hypothesis in (\ref{eq:hyp1}) when $G > 2$, we can use the following test statistic:
\be
BF^{PL}_{ij}(\uPhi) &=& \left(1 + \eta_{ij} \right)^{-m/2} \left\{ \frac{  1 + \frac{m f^{PL}_{max}}{(1 + \eta_{ij})(n-m-(G-1))}}{ 1 + \frac{m f^{PL}_{max}}{(n-m-(G-1))}  } \right\}^{-(n-1)/2}, \label{eq:BFmax}
\ee
where we replace the $(12)$ subscript with the $(ij)$ subscript and refer to it as the Bayes factor (BF) in favor of the alternative for comparing group $i$ and $j$. Additionally, the $(ij)$ subscript denotes the pair with the highest $f^{PL}_{lk}$ value,
i.e., 
$(i, j) = {\mathrm{argmax}}_{(l, k)\in\mathcal{P}}f_{lk}^{PL}$, and
$f^{PL}_{max} = \max_{(l, k)\in\mathcal{P}} f^{PL}_{lk}$
, which is the maximum over all the pairwise $f^{PL}_{lk}$ (data dependent) statistics computed for a random projection across all pairs defined as 
$f^{PL}_{lk} = \frac{n-m-(G-1)}{(n-G)m} n_{0,lk} (\overline{\uX}_l - \overline{\uX}_k)\trans\uPhi(\uPhi\trans\uS_{p}\uPhi\trans)^{-1}\uPhi\trans(\overline{\uX}_l - \overline{\uX}_k)$, where $n = \sum^{G}_{g=1}n_g$, $n_{0,lk} = (1/n_l + 1/n_k)^{-1}$, and $\eta_{lk} = n_{0,lk}/\tau^{PL}_{lk}$ for all $(l, k)\in\mathcal{P}$.
 We use $\tau^{PL}_{ij}$ to denote the scaling factor for the prior covariance matrix under the alternative and allow $\tau^{PL}_{ij}$ to be different across pairs, with the indices $(i,j)$ denoting the pair with highest $f^{PL}_{lk}$ statistic for a specific pair
 $(l, k)\in\mathcal{P}$.
 This pair can be different across different random projection $\uPhi$. Let 
 $1/n_{0,ij} = 1/n_i + 1/n_j$, which is also based on the pair that yields the highest $f^{PL}_{ij}$ statistic. Note that $\uS_p = \sum^{G}_{g=1} (n_g - 1)\uS_{p,g}/(n_1+\cdots +n_{G} - G)$ is the pooled (across all groups) sample covariance and $\uS_{p,g}$ is the group $g$ sample covariance matrix. It is important to note that under the data generating model, $\forall(l,k) \in \mathcal{P}, f^{PL}_{lk} \stackrel{id}{\sim} \uF_{m, n - m- (G-1)}$ (identically but not independently distributed) when $H_{0}$ is true. The lack of independence renders the derivation of the null distribution (or quantiles of the null distribution) of $f^{PL}_{max}$ difficult. We defer the discussion about the choice of $m$ and $\tau^{PL}_{ij}$ to later. 

%
\subsection{Bayes factor-based on paired covariance matrix ($BF^{PR}_{}$)} \label{sec:testid}
The Bayes factor proposed in \eqref{eq:BFmax} relies on a pooled single covariance matrix $\uS_p$ based on the assumption that the covariance matrices across all groups are identical. The assumption of common covariance matrix across groups can reveal very useful as it allows borrowing information across groups to obtain a more precise estimate of the common covariance matrix $\uSigma$, especially in small sample settings. However, it can also be detrimental if grossly wrong. We relax that assumption by instead using a pooled pairwise covariance matrix, which is based on a less stringent assumption than assuming an overall common covariance matrix. Using a similar argument as above, we then get the following test statistic for a single random projection:
\be
BF^{PR}_{ij}(\uPhi) &=& \left(1 + \eta_{ij} \right)^{-m/2} \left\{ \frac{1 + \frac{m f^{PR}_{max}}{(1 + \eta_{ij})(n_{i}+n_{j}-m-1)}}{ 1 + \frac{m f^{PR}_{max}}{(n_{i}+n_{j}-m-1)}  } \right\}^{-(n_{i}+n_{j}-1)/2}, \label{eq:BFmaxij}
\ee
where
$
f^{PR}_{max} = \max_{(l, k)\in\mathcal{P}}f^{PR}_{lk}
$ and the indices pair $(i,j)$ refers to the pair with the highest $f^{PR}_{lk}$ statistic across all $(l, k)\in\mathcal{P}$.
Here $m < \min_{(l, k)\in\mathcal{P}}(n_{l}+n_{k} - 2) << p$,
$f^{PR}_{lk}  = \frac{(n_{l}+n_{k} - m - 1)}{(n_{l}+n_{k} -2)m} n_{0,lk} (\overline{\uX}_l - \overline{\uX}_k)\trans\uPhi (\uPhi\trans\uS_{p,lk}\uPhi)^{-1}\uPhi\trans(\overline{\uX}_l - \overline{\uX}_k)$,
$1/n_{0,lk} = 1/n_l + 1/n_k$, $\uS_{p,lk} =\left\{ (n_{l}-1)\uS_{p,l} + (n_{k}-1)\uS_{p,k} \right\}/(n_{l} + n_{k} -2)$ is the pooled sample covariance matrix for the group $l$ and $k$, 
and $\uS_{p,l}$ is the group $l$ sample covariance matrix.
Finally, $\eta_{lk} = n_{0,lk} /\tau^{PR}_{lk}$, where we allow $\tau^{PR}_{lk}$, the prior scaling factor for $\udelta_{lk}$ under the alternative, to be different across pairs. This allows for an additional flexibility in the prior under the alternative.  
Note that $\forall (l,k) \in \mathcal{P}, \; f^{PR}_{lk} \stackrel{id}{\sim} \uF_{m, n_{l}+n_{k} - m- 1}$ (identically but not independently distributed). Here, the lack of independence also renders the derivation of the distribution of $f^{PR}_{max}$ difficult under $H_0$. 

\subsection{Ensemble test} \label{sec:testens}
\begin{figure}
  \centering
\subfloat[10 groups (45 pairs) ]{\includegraphics[scale=0.35]{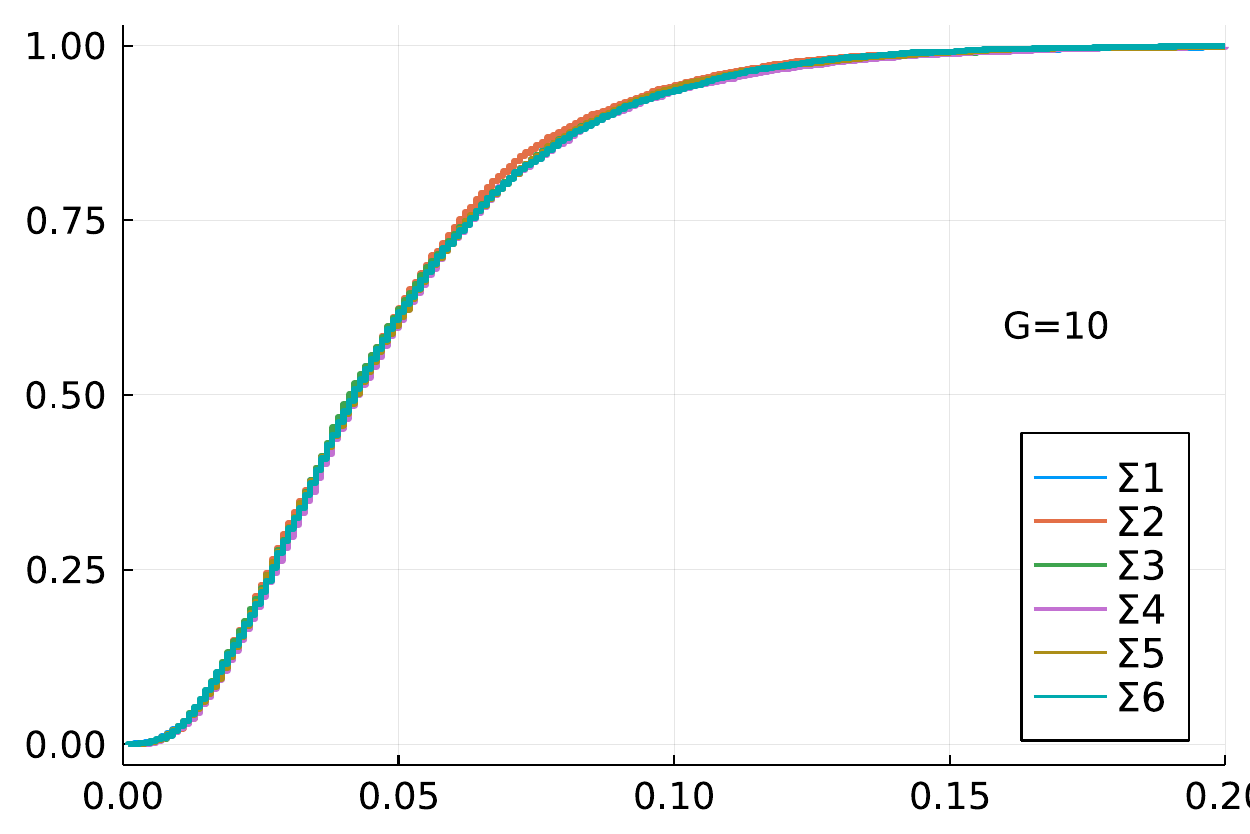}}
\subfloat[3 groups (3  pairs )]{ \includegraphics[scale=0.35]{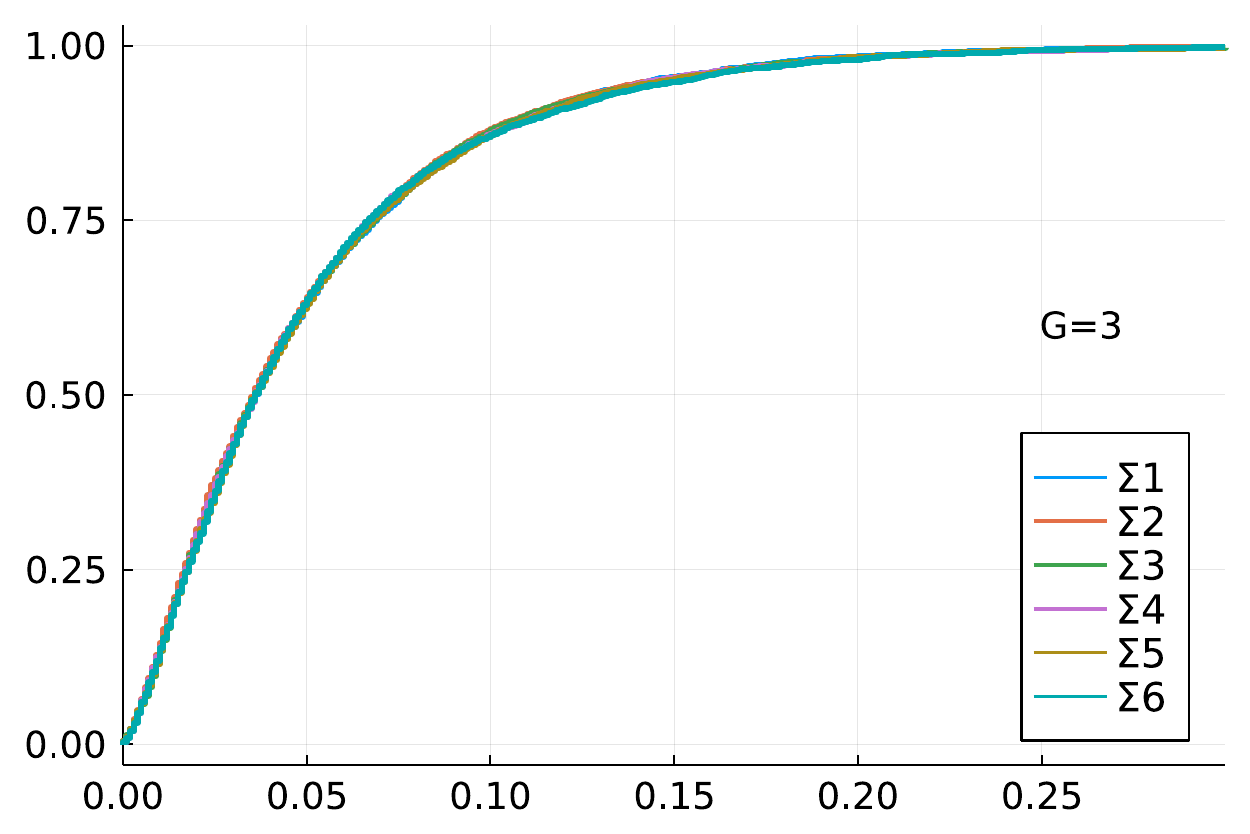}}
\caption{Empirical distribution of the test statistics $\widetilde{\psi}^{PR}$ under the NULL hypothesis of common mean vector and common covariance matrix across groups. This simulation is based on n = 1000 samples and $N = 1000$ random projections using the dense projection matrix.}
  \label{fig:BFh0}
\end{figure}

Based on the BF statistics in (\ref{eq:BFmax}) and (\ref{eq:BFmaxij}), we will decide in favor of the alternative if the $BF$s exceeds a chosen evidence thresholds $\ugamma^{PL}_{ij}$ and $\ugamma^{PR}_{ij}$. 
The ranges of thresholds for Bayes Factors and their interpretation are provided in \cite{kass1995bayes}. We choose to select the evidence thresholds $\ugamma^{PL}_{ij}$ and $\ugamma^{PL}_{ij}$ for our Bayes Factors so to parallel frequentist tests \cite{johnson2013uniformly}. We provide a way to objectively choose the evidence threshold later. A Bayes Factor computed based on a single RP matrix $\uPhi$ can largely dependent on and thus be very sensitive to the choice of that single RP matrix. Instead, we base our final decision on multiple RPs using an ensemble test. Hence, for $N$ randomly chosen RPs matrices, with $N$ sufficiently large, our final test statistic is obtained as
\be
\widetilde{\boldpsi}^{PL}(N) &=& \frac{1}{N}\sum^{N}_{u=1} \bone\{ BF^{PL}_{A_u}(\uPhi_u) \geq \gamma^{PL}_{A_u}\}; \label{eq:BFensblPL}\\
\widetilde{\boldpsi}^{PR}(N) &=& \frac{1}{N}\sum^{N}_{u=1} \bone\{ BF^{PR}_{B_u}(\uPhi_u) \geq \gamma^{PR}_{B_u}\}  \label{eq:BFensblPR}
\ee
where $A_u$ and $B_u$ represent the pair of indices $(i,j)$ on which the Bayes factor is computed for the $u^{th}$ randomly projected version of the original data set; $\bone\{C\}$ is the indicator function which equals $1$ if $C$ is true and zero otherwise.
For the test statistics in \eqref{eq:BFensblPL} and \eqref{eq:BFensblPR}, large values of $\widetilde{\boldpsi}^{PL}_{N}$ and $\widetilde{\boldpsi}^{PR}_{N}$ close to one will tend to  favor the alternatives.
Conversely, lower values of these test statistics will instead favor the NULL hypothesis $H_0$ of no difference in these group mean vectors. Formally, we will make our final decision based on both test statistics using the following rule
\be
 \left \{
       \begin{array}{llll}
       \mbox{Reject}~ H_{0}, & \mbox{if} ~~ \widetilde{\boldpsi}^{PL}(N) > \widetilde{\boldpsi}^{PL}_{0, \alpha},  \\
       \mbox{Accept}~ H_{0}, & \mbox{otherwise}, 
       \end{array}
       \right. \; \mbox{or}\;
  \left \{
       \begin{array}{llll}
       \mbox{Reject}~ H_{0}, & \mbox{if} ~~ \widetilde{\boldpsi}^{PR}(N) > \widetilde{\boldpsi}^{PR}_{0, \alpha},  \\
       \mbox{Accept}~ H_{0}, & \mbox{otherwise}, \label{eq:eqBFfin}
       \end{array}
       \right.     
\ee
where $\widetilde{\boldpsi}^{PL}_{0, \alpha}$ and $\widetilde{\boldpsi}^{PR}_{0, \alpha}$ are  cut-off values for the test statistics $\widetilde{\boldpsi}^{PL}(N)$ and $\widetilde{\boldpsi}^{PL}(N)$, respectively. In the frequentist hypothesis testing scenario, $\widetilde{\boldpsi}^{PL}_{0, \alpha}$ and $\widetilde{\boldpsi}^{PR}_{0, \alpha}$ are selected to achieve a given test size or Type I error rate $\alpha > 0$, commonly selected to be small, say $\alpha  = 0.01, 0.05$, when $H_0$ is true. In essence, $\widetilde{\boldpsi}^{PL}_{0, \alpha}$ and $\widetilde{\boldpsi}^{PR}_{0, \alpha}$ represent the upper $\alpha$ percentiles of the NULL distribution of the test statistics in \eqref{eq:BFensblPR} and \eqref{eq:BFensblPL}, respectively, and will be selected for a chosen Type I error rate $\alpha$ such that the power of our test is comparable to a frequentist test with the same specify Type I error rate.  
Unfortunately, the NULL distribution, distribution of our test statistics when $H_0$ is true and $\udelta_{ij} = \bzero\; \forall(i,j) \in \mathcal{P}$, is difficult to derive analytically. However, under the assumed data generating model, it can be cheaply approximated. Additionally, the NULL distribution of the test statistics is invariant under an arbitrary common mean vector and common (unknown) covariance matrix $\uSigma$. Figure~\ref{fig:BFh0} shows an empirical evidence that the distribution of both test statistics is invariant under the NULL hypothesis of common vector mean and covariance matrix across independent groups. 

We formalize that empirical result later on in the Section~\ref{sec:theori}. 
We also show that the proposed tests are unbiased and their power converges to $1$ with increasing sample size under a sequence of local alternatives (see Section~\ref{sec:theori}).  

\subsection{Choices of $m$, $\tau^{*}_{ij}$, and $\ugamma^{*}_{ij}$} \label{sec:testmtaugam}
\begin{figure}
  \centering
\subfloat[10 groups (45 pairs) ]{\includegraphics[scale=0.35]{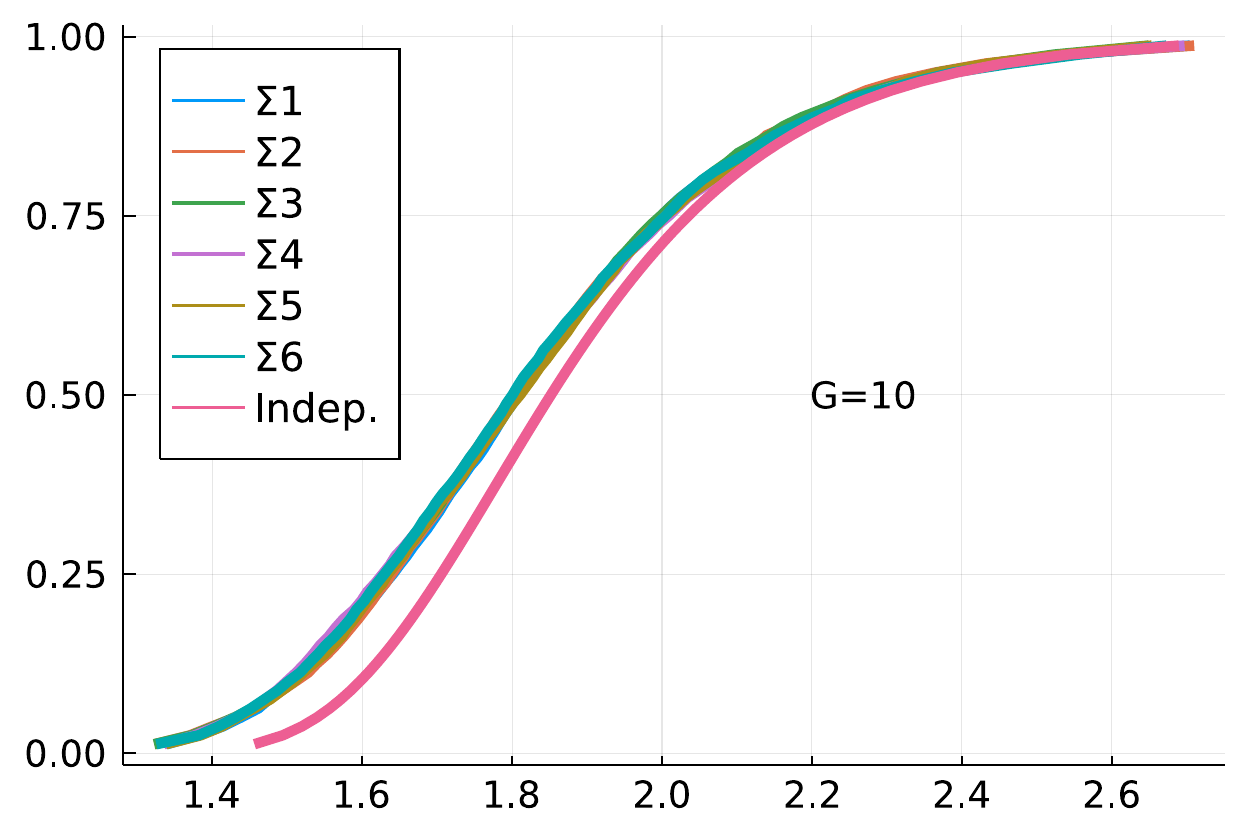}}
\subfloat[3 groups (3  pairs )]{ \includegraphics[scale=0.35]{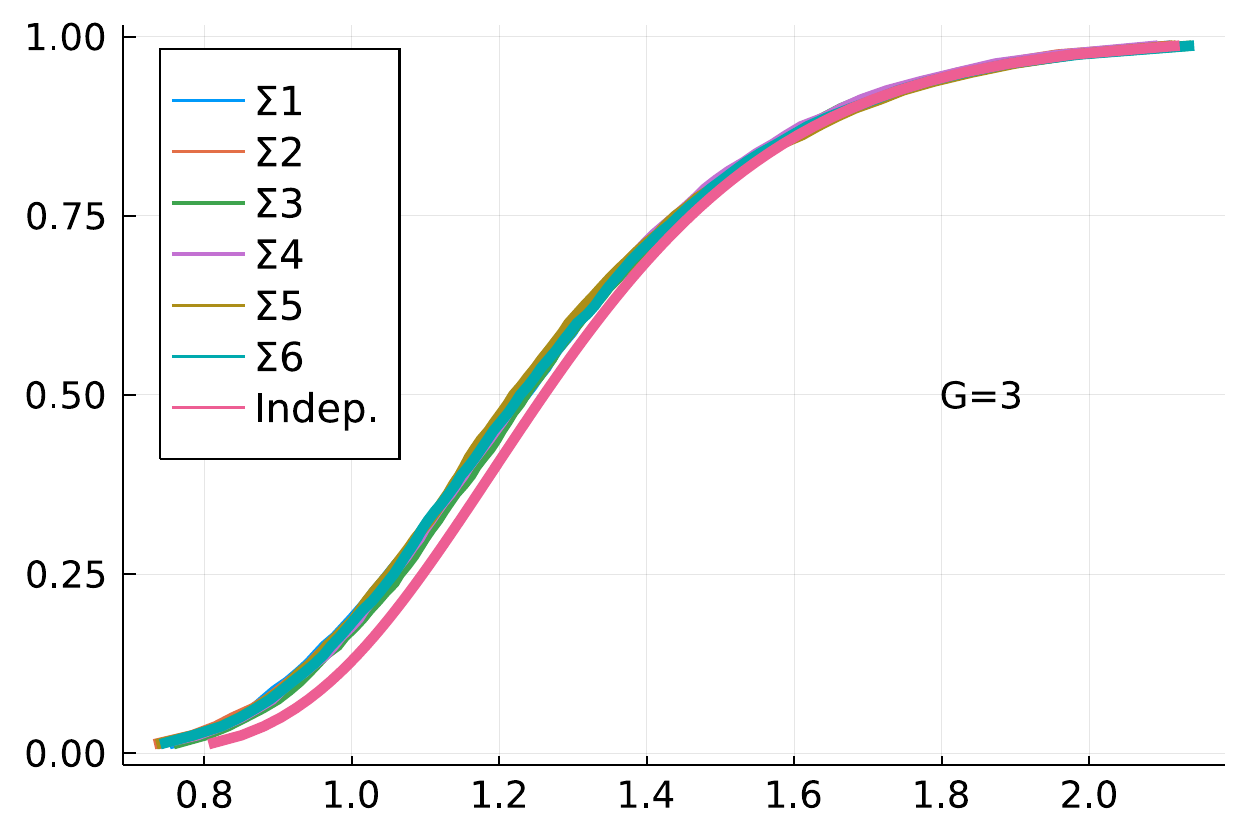}}
  \label{fig:fig2}
\caption{Plots of the empirical distribution of the maximum of identically but correlated $F$ distributed random variables assuming various covariance structure. We also add the case where these $F$ random variables are independent.}
\label{fig:fmaxquant}
\end{figure}

In this section, we will use $\tau^{*}_{ij}$ to simply refer to $\tau^{PL}_{ij}$ or $\tau^{PR}_{ij}$ depending on what BF we are referring to. Similarly, we will use $\ugamma^{*}_{ij}$ to refer to either $\tau^{PL}_{ij}$ or $\tau^{PR}_{ij}$. 
Here, we use $(i, j)$ to denote the $(l, k)$ pair in $\mathcal{P}$ that gives rise to the largest $f_{lk}^{PL}$ or $f_{lk}^{PR}$ statistic, namely, $(i, j) = \mathrm{argmax}_{(l, k)\in\mathcal{P}}f_{lk}^{PL}$ or $(i, j) = \mathrm{argmax}_{(l, k)\in\mathcal{P}}f_{lk}^{PR}$. 
We obtain values for $m$, $\tau^{*}_{ij}$, and $\ugamma^{*}$ for both $BF^{PL}_{ij}$ or $BF^{PR}_{ij}$ using the idea of restricted most powerful Bayesian test (RMPBT) proposed by \cite{GoddardJohnson,Goddard}. To find the RMPBT, we to choose the parameters of the prior distribution under the alternative that maximize the probability of rejecting the NULL under all possible parameters of the data generating model. Namely, for a Bayes Factor in favor of the alternative computed as in \eqref{eq:BFmax} or \eqref{eq:BFmaxij} for testing our hypothesis, we will select $\tau^{*}_{ij}$ so that for a given evidence threshold $\ugamma^{*}_{ij} > 0$ and any other $\tau^{*}_{ij,2}$ ($\tau^{*}_{ij,2} \neq \tau^{*}_{ij}$) associated with a second alternative, we have
$$Pr\{BF^{*}_{ij}(\tau^{*}_{ij}) \geq  \ugamma^{*}_{ij}\} \geq Pr\{BF^{*}_{ij}(\tau^{*}_{ij,2}) \geq \ugamma^{*}_{ij} \},$$
for two different choices of the prior parameters under alternative 1 and alternative 2.
This is equivalent to choosing $\tau^{*}_{ij}$ so that $Pr\{f^{*}_{max} > f^{*}_{max,\;0}(\tau^{*}_{ij},\ugamma^{*}_{ij})\}$ is maximized, which occurs when $f^{*}_{max,\;0}(\tau^{*}_{ij},\ugamma^{*}_{ij})$ is minimized over all possible values of $\tau^{*}_{ij}$ and $\ugamma^{*}_{ij}$. Thus, 
\begin{align*}
    f^{PL}_{max,\;0}(\tau^{PL}_{ij},\ugamma^{PL}_{ij}) =\frac{1+\eta_{ij}}{\eta_{ij}} \left\{ \frac{(N - m -G+1)C^{PL}_{ij}}{m(1 - C^{PL}_{ij})} \right\},\quad
    C^{PL}_{ij} = \frac{1+\eta_{ij}}{\eta_{ij}}\left\{ 1 - \{\ugamma^{PL}_{ij}(1+\eta_{ij})^{m/2} \}^{-2/(N-1)}  \right\},\quad\eta_{ij} = n_{0,ij}/\tau
\end{align*}
for Bayes Factor in ~\eqref{eq:BFmax} and
\begin{align*}
    f^{PR}_{max,\;0}(\tau^{PR}_{ij},\ugamma^{PR}_{ij}) =\frac{1+\eta_{ij}}{\eta_{ij}} \left\{ \frac{(n_i+n_j - m -G+1)C^{PR}_{ij}}{m(1 - C^{PR}_{ij})} \right\},\quad
    C^{PR}_{ij} = \frac{1+\eta_{ij}}{\eta_{ij}}\left\{ 1 - \{\ugamma^{PR}_{ij}(1+\eta_{ij})^{m/2} \}^{-2/(n_i+n_j-1)} \right\},\quad\eta_{ij} = n_{0,ij}/\tau
\end{align*}
for Bayes Factor in ~\eqref{eq:BFmaxij}.
Recalling that the statistics $f^{PL}_{ij} \stackrel{id}{\sim} \uF_{m, N - m- (G-1)} $ and $f^{PR}_{ij} \stackrel{id}{\sim} \uF_{m, n_i+n_j - m- 1}$ respectively when $H_0$ is true, we can select $f^{PL}_{max,\;0}$ and $f^{PR}_{max,\;0}$ so that our test has the same size as an equivalent frequentist test. Namely, for a significance level $\alpha$, we will select $f^{PL}_{max,\;0}(\tau^{PL}_{ij},\ugamma^{PL}_{ij})$ and $f^{PR}_{max,\;0}(\tau^{PR}_{ij},\ugamma^{PR}_{ij})$ so that 
$Pr\{f^{PL}_{max} > f^{PL}_{max,\;0}(\tau^{PL}_{ij},\ugamma^{PL}_{ij}, \alpha)\} = \alpha$ and $Pr\{f^{PR}_{max} > f^{PR}_{max,\;0}(\tau^{PR}_{ij},\ugamma^{PR}_{ij}, \alpha)\} = \alpha$ when $H_{0}$ is true respectively. However, obtaining the upper $\alpha$ percentile of the distributions of $f^{PL}_{max}$ and $f^{PR}_{max}$ is a difficult task. Using a monte carlo step would lead to a significant increase in computation. Under the assumption of common group covariance matrices when $H_0$ is true, we have that: 
\be
 f^{PL}_{max,\;0}(\tau^{PL}_{ij},\ugamma^{PL}_{ij}, \alpha) &\approx& \uF_{m, n -m -(G-1)}\{(1 - \alpha)^{1/|\mathcal{P}|}\}, \nonumber\\
 f^{PR}_{max,\;0}(\tau^{PR}_{ij},\ugamma^{PR}_{ij}, \alpha) &\approx& \underset{(i,j)\; \in\; 
 \mathcal{P}}{\mathrm{argmax}}\; \uF_{m, n_{i}+n_{j} -m -1}\{ (1 -\alpha)^{1/|\mathcal{P}|}\}, \nonumber
 \ee
where $\uF_{a, b}(\theta)$ denotes the upper $\theta$ percentile of an $\uF$ distribution with $a$ and $b$ degrees-of-freedom. This approximation seem to work well for small values of $\alpha$ which we will tend to be concerned with. We plot the exact and the estimated quantile for $f^{PL}_{max,\;0}$ and $f^{PR}_{max,\;0}$ with the case of independence added for comparison (see Figure \ref{fig:fmaxquant}).

We can use that fact to obtain an approximate value of $m$ in both cases as:
\be
 \underset{m\; \in\; (1, n-G)}{\mathrm{arg\max}}\; F_{m, n -m -(G-1)}\{1-(1 - \alpha)^{1/|\mathcal{P}|}\}\;\; \mbox{or}\;\; \underset{m}{\mathrm{\min}} \left\{m:\forall\;(i,j) \in\mathcal{P}\;,  \underset{m\; \in \;(1, n_i+n_{j}-2)}{\mathrm{arg\min}}\; F_{m, n_i+n_j -m -1}\{1 - (1 - \alpha)^{1/|\mathcal{P}|}\} \right\}, \label{eq:mval} 
 \ee
for the BF in (\ref{eq:BFmax}) and (\ref{eq:BFmaxij}) respectively, where $F_{a, b}(\theta)$ is the upper $\theta \in (0, 0.25)$ percentile of a $F$ distribution with $a$ and $b$ degrees of freedom. We discuss the choice of $\alpha$ shortly.
Next, given $n_1,n_2, \cdots, n_{G}$ and $m$, we can reliably approximate the upper quantiles of $f^{PL}_{max}$ and $f^{PR}_{max}$ under the assumption of common covariance matrix using quantiles of a $\uF$ distribution, which provides significant savings in computation time (see Figure~\ref{fig:fmaxquant}). Next we obtain $\tau$ for the pair $(i, j)$ that yield the maximum $f^{PL}_{max}$ and $f^{PR}_{max}$: For a significance level $\alpha$,
\be
\tau^{PL}_{ij} = \frac{n_{0,ij}}{\uF_{m, n -m -(G-1)}\{1-(1-\alpha)^{1/|\mathcal{P}|}\} - 1}\; \mbox{and}\; \tau^{PR}_{ij} = \frac{n_{0,ij}}{\left[\underset{(i,j)\; \in\; \mathcal{P}}{\mathrm{argmin}}\; \uF_{m, n_{i}+n_{j} -m -1}\{1-(1-\alpha)^{1/|\mathcal{P}|}\}\right] - 1} , \label{eq:tau0}
\ee
where $1/n_{0,ij} = 1/n_{i} + 1/n_{j}$. 
Subsequently, we obtain the threshold for each Bayes factor
respectively as
\be
\ugamma^{PL}_{ij} &=& \left\{ 1 + \eta^{PL}_{ij}\right\}^{-m/2}\left\{ 1 - \frac{\eta^{PL}_{ij}}{1 + \eta^{PL}_{ij}} C^{PL}_{ij} \right\}^{-(n -1)/2}, \nonumber \\
\ugamma^{PR}_{ij} &=& \left\{ 1 + \eta^{PR}_{i_1j_1}\right\}^{-m/2}\left\{ 1 - \frac{\eta^{PR}_{i_1j_1}}{1 + \eta^{PR}_{ij}} C^{PR}_{ij} \right\}^{-(n_i+n_j -1)/2}, \nonumber
\ee
where $\ugamma^{PF}_{ij}$ and $\ugamma^{PR}_{ij}$ denotes the evidence threshold values respectively for $\text{BF}_{10}^{PL}$ and $\text{BF}^{PR}_{10}$, $\eta^{PL}_{ij} = n_{0,ij}/ \tau^{PL}_{ij}$, $\eta^{PR}_{ij} = n_{0,ij}/ \tau^{PR}_{ij}$,
and 
\be
\frac{C^{PL}_{ij}}{1 -C^{PL}_{ij}} &=& \frac{ \eta^{PL}_{ij} m}{(1+ \eta^{PL}_{ij})(n - m -(G-1)}f^{PL}_{max,\;0}(\alpha), \nonumber \\
\frac{C^{PR}_{i_1 j_1}}{1 - C^{PR}_{i_1j_1}} &=& \frac{ \eta^{PR}_{i_1 j_1} m}{(1+ \eta^{PR}_{i_1 j_1})(n_{i_1} +n_{j_1} - m -1)}f^{PR}_{max,\;0}(\alpha). \nonumber 
\ee
We note both test statistics could result in different pairs although here we use the same indices for both test statistics. 

Finally, the RP matrices are chosen orthogonal matrices so that for a given RP matrix $\uPhi \in \mathbb{R}^{p \times m}$,  $\uPhi\trans\uPhi = \uI_{m}$. We use the sparse and dense version of the RP matrices proposed by \cite{srivastava2014raptt}. Additionally, the normalization step can be completely skipped based on the results of the QR factorization of a matrix (see \cite{zoh2018powerful} for more details).

\section{Theoretical justifications and Simulation Results} \label{sec:theori}
Here we adopt a slightly different notation to make clear that the quantities we are referring to are dependent on the sample sizes $n_1, \cdots, n_{G}$. Also, let $n_{\min} = \min\{n_1,\cdots, n_{G}\}$, $n_{\max} = \max\{n_1,\cdots, n_{G}\}$, $n = \sum^{G}_{g=1}n_g$, $n_{ij} = n_i + n_j$, and $n_{ij,\min} = \min_{(i,j) \in \mathcal{P}} n_{ij}$. 

\subsection{Theoretical Justifications} 
We assume the following conditions.
\begin{description}
  \item[Assumption 1] $n_g /\sum_{g=1}^{G}n_g \rightarrow \theta_g \in (0, 1)$ for $g=1,\cdots, G$. 
  \item[Assumption 2] $G / n_{min} \rightarrow 0$ as $\min\{ n_1, \cdots, n_G \} \rightarrow \infty$ (G is fixed as a function of $n_{\min}$)
\end{description}

\subsubsection{Consistency of $BF_{}^{PL}$}
\begin{theorem}\label{Thrm1}
Suppose that $X_{ig} = \umu_i + \uepsilon_{ig}$, where $\uepsilon_{ig} \sim \MVN_{p}(\bzero, \uSigma)$, for $g = 1, \cdots, G$ independent groups and $p >> \max\{n_1, \cdots, n_{G}\}$, where $n_1, \cdots, n_{G}$ are the respective sample sizes. 
\begin{enumerate}
    \item If $m \in (1, n - G)$ and $\tau^{PL}_{ij}$ are both fixed and constant functions of sample sizes, 
    then $\log(BF^{PL}_{ij}) \rightarrow -\infty$ under $H_0$ and $\log(BF^{PL}_{ij}) \rightarrow \infty$ under $H_1$ as $n_{min} \rightarrow \infty$. 
    \item If 
    $m_n$
    and $\tau^{PL}_{ij}$ are selected according to our construction in Section~\ref{sec:testmtaugam} 
    and $\alpha < 0.25$, then:
    \begin{itemize}
        \item[(a)] Under $H_0$, $\log(BF^{PL}_{ij}) = \mathcal{O}_{p}(1)$;
        \item[(b)] 
        If the sequence of alternatives $(H_{1,n})_{n = 1}^\infty$ and the projection matrix $\mathbf{\Phi}$ satisfy $\|\mathbf{\Phi}\trans\udelta_{lk}\|_2\to\infty$ and $\|\bSigma\|_2 = O(1)$, then $\log(BF^{PL}_{lk}) \rightarrow \infty$ in probability under $H_{1,n}$. 
    \end{itemize} 
\end{enumerate}
\end{theorem}

\begin{proof}[\textbf{\upshape Proof:}]
Our proof uses similar argument to that of \cite{zoh2018powerful}. 
Recall that $f_{\max}^{PL} = \max_{(l, k) \in\mathcal{P}}f_{lk}^{PL}$ and $(i, j) = \mathrm{argmax}_{(l, k)\in\mathcal{P}}f_{lk}^{PL}$. Namely, $(i, j)$ is a random vector whose distribution depends on the joint distribution of the statistics $(f_{lk}^{PL}:(l, k)\in\mathcal{P})$. The randomness on $(i, j)$ causes complication on the distribution of the aggregated Bayes factor $BF_{ij}^{PL}(\mathbf{\Phi})$. 
Therefore, instead of directly working on $f_{\max}^{PL}$ and $BF_{ij}^{PL}(\mathbf{\Phi})$, for any $(l, k)\in\mathcal{P}$, we define 
\[
\widetilde{BF}_{lk}^{PL}(\mathbf{\Phi}) = \left(1 + \eta^{PL}_{lk} \right)^{-m/2} \left\{ 1 - \frac{\eta^{PL}_{lk}}{(1 + \eta^{PL}_{lk})}\frac{ m f_{lk}^{PL}}{ m f_{lk}^{PL}  + n - m - (G-1)} \right\}^{-(n-1)/2}.
\]
It follows directly that $BF_{ij}^{PL}(\mathbf{\Phi}) = \widetilde{BF}_{ij}^{PL}(\mathbf{\Phi})$ and $\min_{(l,k)\in\mathcal{P}}\widetilde{BF}_{lk}^{PL}(\mathbf{\Phi})\leq BF_{ij}^{PL}(\mathbf{\Phi})\leq \max_{(l,k)\in\mathcal{P}}\widetilde{BF}_{lk}^{PL}(\mathbf{\Phi})$. Therefore, for the remaining proof, it is sufficient to focus on any fixed $(l, k)\in\mathcal{P}$ and $\widetilde{BF}_{lk}^{PL}(\mathbf{\Phi})$. 
\begin{description}
\item[Part(1)]
For $1 < m < n - G$ and 
$(l, k)\in\mathcal{P}$,
we integrate out the parameters with respect to the conjugate priors to obtain the Bayes Factor in favor of the alternative as
\be
\widetilde{BF}^{PL}_{lk}(\uPhi) &=&
\left(1 + \eta^{PL}_{lk} \right)^{-m/2} \left\{ 1 - \frac{\eta^{PL}_{lk}}{(1 + \eta^{PL}_{lk})}\frac{ m f_{lk}^{PL}}{ m f_{lk}^{PL}  + n - m - (G-1)} \right\}^{-(n-1)/2} 
\nonumber,
\ee
where
\bse
f_{lk}^{PL}  = \frac{n - m - (G-1)}{(n-G)m} n_{0,lk}(\overline{\uX}_l - \overline{\uX}_k)\trans \uPhi(\uPhi\trans \uS\uPhi )^{-1}\uPhi\trans (\overline{\uX}_l - \overline{\uX}_k)
\ese
and $$ \uS = \frac{1}{n - G} \sum^{G}_{g=1}(n_g -1)\uS_{g}\; \text{and} \;\; \uS_g = \frac{1}{n_g - 1}\sum^{n_g}_{i=1}(\uX_{ig} - \overline{\uX}_{g})(\uX_{ig} - \overline{\uX}_{g})\trans.$$
Recall that $1/n_{0,lk} = 1/n_l + 1/n_k$, $\eta^{PL}_{lk} = n_{0,lk}/\tau^{PL}_{0,lk}$, and $n_{\min} = \min\{n_1, \cdots, n_G\}$.
Since $\tau^{PL}_{lk}$ is fixed, $\eta^{PL}_{lk} \rightarrow \infty$ as $n_{\min} \to \infty$.
For a randomly chosen projection matrix $\uPhi$, under $H_{0}$, $f^{PL}_{lk}  \sim F_{m, n-m-(G-1)}$ with $m$ and $n - m -(G-1)$ degrees of freedom.
Thus, $f^{PL}_{lk} = O_{p}(1)$ and $f^{PL}_{\max} = \max_{(l,k) \in \mathcal{P}}\{ f_{lk}^{PL}\}$.
Also, from well-known properties of the $F$ distribution, we have that
\begin{align*}
U_{lk} & = \frac{m f_{lk}^{PL}/(n-m-(G-1) )}{\{m f_{lk}^{PL}/(n-m- (G-1) ) + 1\}} 
  =\frac{m f_{lk}^{PL}}{( m f_{lk}^{PL} +n-m- (G-1))} \sim \Beta\{m/2, (n-m-(G-1))/2 \},
\end{align*}
for each {$(l, k)\in\mathcal{P}$}, where $\Beta(a, b)$ denotes a Beta distribution.
Therefore, $\{\eta^{PL}_{lk}/(1 + \eta^{PL}_{lk})\} U_{lk} = o_{p}(1)$ by Markov's inequality because $\mathbb{E}(U_{lk}) \to 0$ with $m/n\to 0$.
(I found the proof here slightly not rigorous so I did a little asymptotic analysis) Since $\log(1 - x) = O(x)$ as $x\to 0$, then $\log(1 - X) = O_p(X)$ if $X = o_p(1)$, and hence,
\begin{align*}
    \frac{(n - 1)}{2}\log\left\{1 - \frac{\eta_{lk}^{PL}}{1 + \eta_{lk}^{PL}}U_{lk}\right\}
    & = \frac{(n - 1)}{2}O_p\left\{\frac{\eta_{lk}^{PL}}{1 + \eta_{lk}^{PL}}U_{lk}\right\} = \frac{\eta_{lk}^{PL}}{1 + \eta_{lk}^{PL}}O_p(nU_{lk}) = O_p(1)
\end{align*}
by Markov's inequality because $\mathbb{E}(nU_{lk}) = O(1)$. 
We then get
\bse
-\frac{m}{2}\log(1 + \eta^{PL}_{lk}) -\frac{(n-1)}{2}\log\left\{1 - \frac{\eta_{lk}^{PL}}{1 + \eta_{lk}^{PL}}U_{lk}\right\}
= -\frac{m}{2}\log(1 + \eta^{PL}_{lk}) - O_p(1) \xrightarrow[]{p} -\infty,
\ese
since $\log(1 + \eta^{PL}_{lk}) \rightarrow \infty$ as $n_{\min} \rightarrow \infty$ and $\lim_{n_{\min} \rightarrow \infty} m = m > 0$.
We conclude that $\log\{\widetilde{BF}^{PL}_{lk}(\uPhi)\} \xrightarrow[]{p} -\infty$ under the null hypothesis for all $(l, k)\in\mathcal{P}$.
This result hold for any $(l,k) \in \mathcal{P}$ and we conclude $\log\{BF^{PL}_{ij}(\uPhi)\} \xrightarrow[]{p} -\infty$. 

Under the alternative, 
there exists some $(l, k)\in\mathcal{P}$ such that
$\umu_{l} \neq \umu_{k}$ and $\udelta_{lk} \sim \uN_{p}({\bf 0}, \uSigma /\tau^{PL}_{lk})$.
Then, $f_{lk}^{PL} \mid   \lambda_{lk} \sim F_{m, n-m-(G-1)}(\lambda_{lk})$
with non-centrality $\lambda_{lk} = n_{0,lk}\udelta_{lk}\trans\uPhi(\uPhi\trans \uSigma \uPhi)^{-1}\uPhi\trans \udelta_{lk}$.
Since $\udelta_{lk} \sim \uN_{p}({\bf 0}, \uSigma /\tau_{lk})$, $\lambda_{lk} \sim n_{0,lk}\chi_{m}^{2}/\tau_{lk}$,
where $\chi_{m}^{2}$ denotes a $\chi^{2}$ distribution with m degrees of freedom. The non-centrality parameter depends on $n$ through $n_{0,lk}$.
It can be shown that the unconditional distribution of $f_{lk}^{PL} /(1 + \eta^{PL}_{lk}) \sim F_{m, n-m-(G-1)}$ \citep[see for reference][page 704]{johnson2005bayes}.
If we denote $f_{0,lk}^{PL} =  f_{lk}^{PL} /(1 + \eta^{PL}_{lk})$, since $m$ is fixed and $n - m - (G - 1)\to\infty$, then $mf_{0,lk}^{PL}\overset{\mathcal{L}}{\to}\chi_m^2$ by the definition of $F$-distribution. We have that
\be
\frac{\eta^{PL}_{lk} U_{lk}}{(1+\eta^{PL}_{lk})} = \frac{ m f_{lk}^{0}}{m f_{lk}^{0}(1+\eta^{PL}_{lk})/\eta_{lk}^{PL}+(n-m-(G-1))/\eta_{lk}^{PL}} \nonumber.
\ee
Because $\eta_{lk}^{PL}\to\infty$, then $(1+\eta^{PL}_{lk})/\eta_{lk}^{PL}\to 1$ and by Assumption 1, $(n - m - (G - 1))/\eta_{lk}^{PL}$ also converges to a constant as $n_{\min}\to\infty$. 
This implies that $\eta^{PL}_{lk} U_{lk}/(1+\eta^{PL}_{lk})$ is bounded below in probability. Also, observe that $\eta_{ij}^{PL}/(1 + \eta_{ij}^{PL}) = \eta_{lk}^{PL}/(1 + \eta_{lk}^{PL})\{1 + o_p(1)\}$.
Therefore, by the basic inequality $\log(1 - x)\leq -x$ for all $x < 1$ and the fact that $U_{ij}\geq U_{lk}$ because $U_{lk}$ is increasing with respect to $f_{lk}^{PL}$, we have
\begin{align*}
    \log({BF}_{ij}^{PL})
    & = -\frac{m}{2}\log(1 + \eta_{ij}) - \frac{(n - 1)}{2}\log \left\{1 - \frac{\eta^{PL}_{ij} U_{ij}}{(1+\eta^{PL}_{ij})}\right\}\\
    &\geq -C\sqrt{n} + \frac{(n - 1)}{2}\frac{\eta^{PL}_{ij} U_{ij}}{(1+\eta^{PL}_{ij})}
    \geq \frac{(n - 1)}{2}\left\{\frac{\eta^{PL}_{lk} U_{lk}}{(1+\eta^{PL}_{lk})} - o_p(1)\right\}\overset{p}{\to}\infty
\end{align*}
as $n_{\min}\to\infty$, where $C > 0$ is some constant. 
Hence $\log\left\{BF^{PL}_{ij}(\uPhi) \right \} \xrightarrow[]{p} \infty$ under the alternative. 

\item[Part(2)] First let's show that $m_n/n \rightarrow a \in (0, 1)$. 
We have that for a chosen $G$ and samples sizes $n_1, \cdots, n_{G}$,  $F_{\alpha, m, n-m_n-(G-1)}$ is convex over the range of possible values of $m_n \in (1, n-G)$. For large values of $m_n$ and $n - m_n$, we have $F_{\alpha, m_n, n-m_n-(G-1)}$ is convex suggesting that $m_n$ and $n-m_n$ diverge. 

{\color{black}
We prove that $m_n\to\infty$ by contradiction, and the proof of the claim that $n - m_n\to\infty$ is similar by taking the reciprocal of the $F$ distribution. Namely, there exists a subsequence $(n_k)_{k = 1}^\infty$ such that $m_{n_k}\to \bar{m}$ for some fixed $\bar{m}\in\mathbb{N}_+$ as $k\to\infty$. Since $m_n$'s are integers, it follows that $m_{n_k} = \bar{m}$ for sufficiently large $k$. By the properties of the $F$ distribution, we have $F_{m_{n_k}, n_k - m_{n_k} - (G - 1)} \overset{\mathcal{L}}{\to} \chi^2_{\bar{m}}/\bar{m}$, where $\chi^2_{\bar{m}}$ is the chi-squared distribution with degree of freedom $\bar{m}$. By the convergence of the quantile function, this implies that $F_{\alpha, m_{n_k}, n_k - m_{n_k} - (G - 1)}\to \chi_{\alpha,\bar{m}}^2/\bar{m}$, where $\chi_{\alpha,\bar{m}}^2$ is the upper $\alpha$ quantile of $\chi_{\bar{m}}^2$. Since $\alpha < 0.25$, by the Berry-Esseen bound, we have that $\chi_{\alpha,\bar{m}}^2/\bar{m} > 1$ for any fixed $\bar{m}$, implying that $F_{\alpha, m_{n_k}, n_k - m_{n_k} - (G - 1)} > 1$ for sufficiently large $k$. On the other hand, if $(m_n^*)_{n = 1}^\infty$ is a sequence with $m^*_n/\sqrt{n}\to 1$ as $n\to\infty$, then by the central limit theorem and the weak law of large numbers, $\sqrt{m_n^*}(F_{m_n^*,n - m_n^* - (G - 1)} - 1)\overset{\mathcal{L}}{\to}\mathrm{N}(0, 2)$, thereby implying that $F_{\alpha, m_n^*, n - m_n^* - (G - 1)}\to 1$ by the convergence of the quantile. This further shows that $F_{\alpha, m_{n_k}^*, n_k - m_{n_k}^* - (G - 1)} < F_{\alpha, m_{n_k}, n_k - m_{n_k} - (G - 1)}$ for sufficiently large $k$. Hence, $m_{n_k}$ cannot be the minimizer of $F_{\alpha, m, n_k - m - (G - 1)}$ w.r.t. $m$ because $m_{n_k}^*$ gives a smaller value of the quantile. This contradicts with the definition of $m_{n_k} = \arg\min_mF_{\alpha, m, n_k - m - (G - 1)}$, and thus, we have $\liminf_{n\to\infty}m_n = \infty$.
}
We have that for large $m_n$ and $n-m_n$, then $F_{\alpha, m_n, n-m_n-(G-1)} \approx \mu_n + \sigma_n\Phi^{-1}(\alpha)$, where $\mu_n = \frac{n-m_n-(G-1)}{n-m_n-(G-1) - 2} > 1$ and $\sigma^2_n = \frac{2(n-m_n-(G-1))^2 (n-(G-1)-2)}{m_n(n-m_n-(G-1))(n-m_n-(G-1)-2)^2}$; $\Phi^{-1}(\alpha)$ is the upper $\alpha$ percentile of the standard normal distribution. Thus the quantile of F distribution is at it minimum if when $\sigma^2_n$ is minimum. Thus, using the result from \cite{zoh2018powerful}, we have that $m_n \approx n/2$ thus $m_n/n \rightarrow 1/2$ as $n_{\min} \rightarrow \infty$. 

For any $(l, k)\in\mathcal{P}$, 
$\eta^{PL}_{lk} = n_{0,lk}/\tau^{PR}_{lk} = F_{1-(1-\alpha)^{1/|\mathcal{P}|},m_n, n-m_n-(G-1)} - 1 \rightarrow 0$, since $\mu_m \rightarrow 1$ and $\sigma^2_n = \mathcal{O}(1/m_n)$ and thus  $F_{1-(1-\alpha)^{1/|\mathcal{P}|},m_n, n-m_n-(G-1)} \rightarrow 1$ converges to $1$ as $n_{\min} \rightarrow \infty$. 
We see that $F_{1-(1-\alpha)^{1/|\mathcal{P}|}, m_n, n-m_n-(G-1)} -1 $ converges to 0 at a slower rate than $1/\sqrt{n}$. We conclude that $n\eta^{PR}_{lk} \rightarrow \infty$ and $m_{n}\eta^{PR}_{lk} \rightarrow \infty$ as $n_{\min} \rightarrow \infty.$

We have the following expression for the $\log$ Bayes factor. 
For any $(l, k)\in\mathcal{P}$,
\be
 \log\{\widetilde{BF}^{PL}_{lk}(\uPhi)\} =\frac{n}{2}\left( 1 - \frac{m_n}{n}\right)\log\left(1 + \eta^{PL}_{lk} \right) - \frac{n}{2}\log\left\{ 1 + \eta^{PL}_{lk}(1-U_{lk}) \right \} +\frac{1}{2}\log \left\{1 - \frac{\eta^{PL}_{lk} U_{lk}}{1+\eta^{PL}_{lk}}\right\},  \nonumber
\ee
where $U_{lk} \sim Beta\left\{ m_n/2,  (n - m_n -(G-1))/2 \right\}$ under $H_0$ for each $1 \leq l < k \leq G$.
{\color{black}
By Taylor's expansion, we have
$\log(1 + \eta_{lk}^{PL}) = \eta_{lk}^{PL} + O\{(\eta_{lk}^{PL})^2\}$ and $\log\{1 + \eta_{lk}^{PL}(1 - U_{lk})\} = \eta_{lk}^{PL}(1 - U_{lk}) + O_p\{(\eta_{lk}^{PL})^2(1 - U_{lk})^2\}$ as $\eta_{lk}^{PL}\to 0$. Then a further computation leads to
\begin{align*}
    &\frac{n}{2}\left( 1 - \frac{m_n}{n}\right)\log\left(1 + \eta^{PL}_{lk} \right) - \frac{n}{2}\log\left\{ 1 + \eta^{PL}_{lk}(1-U_{lk}) \right \}\\
    &\quad = \frac{n}{2}\left( 1 - \frac{m_n}{n}\right)\eta_{lk}^{PL} + O\{(n\eta_{lk}^{PL})^2\} - \frac{n}{2}\eta_{lk}^{PL}(1 - U_{lk}) + O_p\{n(\eta_{lk}^{PL})^2(1 - U_{lk})^2\}\\
    &\quad = \frac{n}{2}\left( 1 - \frac{m_n}{n}\right)\eta_{lk}^{PL} - \frac{n}{2}\eta_{lk}^{PL}\left\{1 - \frac{m_n}{n} + O_p\left(\frac{1}{\sqrt{n}}\right)\right\} + O_p\{n(\eta_{lk}^{PL})^2\}\\
    &\quad = O_p(\sqrt{n}\eta_{lk}^{PL}) + O_p\{n(\eta_{lk}^{PL})^2\}
\end{align*}
because Chebyshev's inequality implies that $U_{lk} = m_n/n + O_p(n^{-1/2})$. Since the $F$-quantile has the approximation
\begin{align*}
F_{1 - (1 - \alpha)^{1/|\mathcal{P}|}, m_n, n - m_n - (G - 1)} 
&\approx \frac{n - m_n - (G - 1)}{n - m_n - (G - 1) - 2} + \sigma_n\Phi^{-1}(1 - (1 - \alpha)^{1/|\mathcal{P}|})\\
& = 1 + O\left\{\frac{1}{n - m_n - (G - 1) - 2}\right\} + O\left(\frac{1}{\sqrt{m_n}}\right) = 1 + O\left(\frac{1}{\sqrt{n}}\right),
\end{align*}
it follows that $\eta_{lk}^{PL} = F_{1 - (1 - \alpha)^{1/|\mathcal{P}|}, m_n, n - m_n - (G - 1)} - 1 = O(n^{-1/2})$. We also remark that the same derivation implies that $\eta_{lk}^{PL}\geq cn^{-1/2}$ for some constant $c > 0$ since $1 - (1 - \alpha)^{1/|\mathcal{P}|} < 1/2$, i.e., $\Phi^{-1}(1 - (1 - \alpha)^{1/|\mathcal{P}|})\neq 0$. 
}
The distribution $ \log\{\widetilde{BF}^{PL}_{lk}(\uPhi) \}$ then depends on that of $U_{lk}$, which
converges to $\lim_{n\to\infty}m_n/n$ in probability (by the Chebyshev's inequality because $\var(U_{lk})\to 0$)
when the null hypothesis is true.
Therefore, under $H_0$, $\log\{\widetilde{BF}^{PL}_{lk}(\uPhi) \} = \mathcal{O}_{p}(1)$ for all $(l, k)\in\mathcal{P}$ and we conclude $\log\{BF^{PL}_{ij}(\uPhi) \} = \mathcal{O}_{p}(1)$.

Under $H_{1,n}$ and $\umu_{k} \neq \umu_{l}$ for a $(l, k) \in \mathcal{P}$, again 
by the fact that $U_{ij}\geq U_{lk}$ 
we have
\begin{align*}
    \log\{BF^{PL}_{ij}(\uPhi) \} &= -\frac{m_n}{2}\log(1 + \eta^{PL}_{ij}) - \frac{(n-1)}{2}\log\left\{ 1 - \frac{\eta^{PL}_{ij} U_{ij}}{1 + \eta^{PL}_{ij}} \right\}\\
    & \geq \left\{\frac{(n - 1)}{2} - \frac{m_n}{2}\right\}\log(1 + \eta^{PL}_{ij}) - \frac{(n - 1)}{2}\log\{1 + \eta_{ij}^{PL}(1 - U_{lk})\}
\end{align*}
{\color{black}
We now argue that $U_{lk}\to 1$ in probability, which is equivalent to showing that $f_{lk}^{PL}\to\infty$. Recall that $f_{lk}^{PL}\sim F_{m_n, n - m_n - (G - 1)}(\lambda_{lk})$ and $\lambda_{lk} = n_{0,lk}\udelta_{lk}\trans\mathbf{\Phi}(\mathbf{\Phi}\trans\bSigma\mathbf{\Phi})^{-1}\mathbf{\Phi}\trans\udelta_{lk}$. Clearly,
\[
\lambda_{lk} \geq n_{0,lk}\|\mathbf{\Phi}\trans\udelta_{lk}\|_2^2\lambda_{\min}\{(\mathbf{\Phi}\trans\bSigma\mathbf{\Phi})^{-1}\}
= \frac{n_{0,lk}\|\mathbf{\Phi}\trans\udelta_{lk}\|_2^2}{\lambda_{\max}\{(\mathbf{\Phi}\trans\bSigma\mathbf{\Phi})}\geq \frac{n_{0,lk}\|\mathbf{\Phi}\trans\udelta_{lk}\|_2^2}{\|\bSigma\|_2} \geq Cn\|\mathbf{\Phi}\trans\udelta_{lk}\|_2^2 
\]
for some constant $C > 0$, where $\lambda_{\min}$ and $\lambda_{\max}$ denote the smallest and the largest eigenvalue of a positive definite matrix, respectively. 
Since $m_n/n\to 1/2$, it follows that $f_{lk}^{PL}\to \infty$ in probability because the denominator in the $F$-distribution converges to $1$ in probability by the weak law of large numbers and the expected value of the numerator has the form $1 + \lambda_{lk}/m_n\to\infty$. This completes the proof for $U_{lk}\to 1$ in probability. 

Also recall that $\eta_{lk}^{PL} \geq cn^{-1/2}$ for some constant $c > 0$ for all $(l,k)\in\mathcal{P}$. Therefore, we conclude that there exists constants $C_1, C_2 > 0$, such that for sufficiently large $n$,
\begin{align*}
    \log\{BF^{PL}_{ij}(\uPhi) \} & \geq \left\{\frac{(n - 1)}{2} - \frac{m_n}{2}\right\}\log(1 + \eta^{PL}_{ij}) - \frac{(n - 1)}{2}\log\{1 + \eta_{ij}^{PL}(1 - U_{lk})\}\\
    &\geq \left\{\frac{(n - 1)}{2} - \frac{m_n}{2}\right\}C_1\eta_{ij}^{PL} - \frac{(n - 1)}{2}C_2\eta_{ij}^{PL}(1 - U_{lk})\\
    & = \eta_{ij}\left[C_1\left\{\frac{(n - 1)}{2} - \frac{m_n}{2}\right\} - C_2\frac{(n - 1)}{2}(1 - U_{lk})\right]\to \infty
\end{align*}
in probability. The proof is thus completed. 
}
\end{description}
\end{proof}

\subsubsection{Consistency of $BF_{}^{PR}$}
\begin{theorem}\label{Thrm2}
Suppose that $X_{ig} = \umu_i + \uepsilon_{ig}$, where $\uepsilon_{ig} \sim \MVN_{p}(\bzero, \uSigma)$, for $g = 1, \cdots, G$ independent groups and $p >> \max\{n_1, \cdots, n_{G}\}$, where $n_1, \cdots, n_{G}$ are the respective sample sizes. 
\begin{enumerate}
    \item If $m \in (1,  n_{ij} - 2)$ and $\tau^{PR}_{ij}$ are fixed and constant function the sample sizes, 
    then $\log(BF^{PR}_{ij}) \rightarrow -\infty$ under $H_0$ and $\log(BF^{PR}_{ij}) \rightarrow \infty$ under $H_1$ as $n_{min} \rightarrow \infty$. 
    \item If $m_n$ and $\tau^{PR}_{ij}$ are selected according to our construction in Section~\ref{sec:testmtaugam}, then 
    \begin{itemize}
     \item[(a)]  $\log(BF^{PR}_{ij}) = \mathcal{O}_{p}(1)$ under $H_0$ 
     \item[(b)] 
     If the sequence of alternatives $(H_{1,n})_{n = 1}^\infty$ and the projection matrix $\mathbf{\Phi}$ satisfy $\|\mathbf{\Phi}\trans\udelta_{lk}\|_2\to\infty$ and $\|\bSigma\|_2 = O(1)$, then $\log(BF^{PL}_{lk}) \rightarrow \infty$ in probability under $H_{1,n}$. 
    \end{itemize}
\end{enumerate}
\end{theorem}
The proof of Theorem \ref{Thrm2} is very similar to Theorem~\ref{Thrm1} and is omitted here. 

\refmark{\bf Remarks:}
We make the following remarks about the both Bayes factors. 
\begin{enumerate}
    \item Theorems~\ref{Thrm1} and ~\ref{Thrm2} show that both Bayes factors we constructed have the behavior of the usual Bayes Factor, i.e, the consistency for a chosen projection dimension $m$, alternative scale parameter $\tau_{ij}$, and evidence threshold $\gamma_{ij}$ all fixed function of the sample sizes. 
    \item However when $m_n$, $\tau^{\star}_{ij}$, $\gamma^{\star}_{ij}$ are selected according to our prescription as in Section~\ref{Thrm2}, both Bayes factors we constructed are bounded in probability under the null hypothesis. But under the sequence of alternative $H_{1,n}$ associated with $\tau^{*}_{ij}$ (dependent on sample sizes), the Bayes Factor converges to $+\infty$ under certain regularity conditions.
\end{enumerate}

\subsubsection{Power of the ensemble test}
\begin{theorem}
Suppose the assumptions of Theorems~\ref{Thrm1} and ~\ref{Thrm2} hold.
Given a collection $\uPhi_1, \cdots, \uPhi_N$ of independent random projections matrices,
where $\uPhi_{i}\trans\uPhi_{i} = \uI$ for all $i = 1, \cdots, N$ ($N$ potential large), then $\lim_{n_{\min} \rightarrow \infty} Pr\{\widetilde{\boldpsi}^{\star}(N) > \widetilde{\boldpsi}^{\star}_{0,\alpha}\}  = 1$ under the sequence $H_{1,n}$ of alternatives,
where $\widetilde{\boldpsi}^{\star}(N)\;(\widetilde{\boldpsi}^{\star}_{0,\alpha})$ denotes either the $\widetilde{\boldpsi}^{PL}(N)\; (\widetilde{\boldpsi}^{PL}_{0,\alpha})$ or $\widetilde{\boldpsi}^{PR}(N)\;(\widetilde{\boldpsi}^{PR}_{0,\alpha})$. 
\end{theorem}
\begin{proof}[\textbf{\upshape Proof:}]
The proof is similar to that of \cite{zoh2018powerful} and the will use the $(\star)$ to denote either tests.  
The power of our test is $Pr\{ \widetilde{\psi}^{\star}(N) > \widetilde{\psi}^{\star}_{0,\alpha}\mid  H_{1,n}\}$.
Henceforth, we make it explicit that $\widetilde{\psi}^{\star}_{0,\alpha}$ depends on $(n_1, \cdots ,n_G)$ and write $\widetilde{\psi}^{\star}_{0,\alpha}(n_1, \cdots, n_G)$ instead.

Given $n_1, \cdots, n_G$ and $\alpha$, we choose $\widetilde{\psi}^{\star}_{0,\alpha}(n_1, \cdots, n_G)$ so that $Pr\{\widetilde{\psi}^{\star}(N) > \widetilde{\psi}^{\star}_{0, \alpha}(n_1, \cdots,n_G)\mid  H_{0}\}  = \alpha$.
Since $ 0 < \widetilde{\psi}^{\star}_{0,\alpha}(n_1, \cdots,n_G) < 1$, for $0 < \alpha < 1$, we have that $Pr\{ \sum^{N}_{i=1}\widetilde{\psi}^{\star}(\uPhi_i) \geq 0 \mid H_{1,n} \} \geq Pr\{ \sum^{N}_{i=1}\widetilde{\psi}^{\star}(\uPhi_i) > N\widetilde{\psi}^{\star}_{0,\alpha}(n_1, \cdots ,n_G) \mid H_{1,n} \} \geq Pr\{ \sum^{N}_{i=1}\widetilde{\psi}^{\star}(\uPhi_i) \geq N \mid H_{1,n} \}$.

We have that $Pr\{\widetilde{\psi}^{\star}(\uPhi_i) =1 \mid H_{1,n}\} \rightarrow 1$ as $n_{\min} \rightarrow \infty$, under the alternative for $i = 1, \cdots, N$. So, $Pr\{ \sum^{N}_{i=1}\widetilde{\psi}^{\star}(\uPhi_i) \geq 0 \mid H_{1,n} \} = 1 - \prod_{i=1}^{N}Pr\{\widetilde{\psi}^{\star}(\uPhi_i) = 0 \mid H_{1,n} \} \rightarrow 1$.
Additionally, $Pr\{ \sum^{N}_{i=1}\widetilde{\psi}^{\star}(\uPhi_i) \geq N \mid H_{1,n} \} = Pr\{ \sum^{N}_{i=1}\widetilde{\psi}^{\star}(\uPhi_i) = N \mid H_{1,n} \} = \prod_{i=1}^{N}Pr\{\widetilde{\psi}^{\star}(\uPhi_i) = 1 \mid H_{1,n}\} \rightarrow 1$ for fixed $N$ as $n_{\min} \rightarrow \infty$. We conclude that $Pr\{ \widetilde{\psi}^{\star}(N) \geq \widetilde{\psi}^{\star}_{0, \alpha}\mid  H_{1,n}\} \rightarrow 1$ as $n_{\min} \rightarrow \infty.$
\end{proof}

\subsection{Simulation}

\subsubsection{Simulation Study design}
We designed a simulation study aiming at investigating the power of the tests proposed in Section~\ref{sec:test}
with respect to a sparse true mean vector under the alternative. The proportion of true elements of $\umu$ that are actually zero are varied along with the covariance matrices.
Thus, we considered two settings for our simulation. In each case, we had two conditions for each choice of the covariance matrix. 
In the first condition, we assumed $p = 200$, $G=3$ and $5$, and $n_g = 50\;\forall\; g =  1, \cdots, G$. Using the approach described above (Section~\ref{sec:test}), we find $m = 43$ for the test based on $BF^{PR}_{}$ for both $G=3$ and $5$. However, for the test based on $BF^{PR}$, we get $m = 65$ and $m = 111$ when $G=3$ and $G=5$, respectively. In the second condition, $p = 1000$, $G=3$ and $5$, and $n_g = 70$, $\forall\; 1 \leq g \leq G$. In this condition, for the test based on $BF^{PR}_{}$, $m = 62$ and for the test based on $BF^{PL}_{}$, $m = 105$ and $175$ for $G=3$ and $G=5$, respectively. We denote the proportion of entries of the vector $\udelta$ that are exactly zero with $p_0$. We chose $p_0 = 0.5, .75, .80, 0.95, 0.99$, and $1.00$ (null hypothesis). In each setting, the values of $\tau^{\star}_{ij}$ and $\gamma^{\star}_{ij}$ were chosen according to our discussion in Section~\ref{sec:testmtaugam} for both tests. We considered two types of random projections matrices, $\uPhi_{1}$(full matrix)  and $\uPhi_2$ (sparse matrix), as previously described \cite{srivastava2014raptt,zoh2018powerful}. Finally, we assumed $\alpha = 0.05$. In each setting, we estimated the power of our tests based on $1000$ random samples and $N = 1000$ independent random projection matrices.

In \textbf{case 1}, only the last group $G$ had a non-zero mean vector $\umu_G$ and all the others groups had vector mean zero under the alternative. In \textbf{case 2}, however, only the last group $G$ had a zero vector mean $\umu_{G}$ under the alternative.
We considered the following choices of covariance matrix $\uSigma=(\sigma_{ij})$:
\begin{enumerate}
  \item $\uSigma_{1} = \uI_{p \times p} $ is the identity matrix.
  \item $\uSigma_{2} $ is a block diagonal matrix, with block $\uB = 0.85\uI_{25 \times 25} + 0.15\uJ_{25 \times 25}$. $\uJ$ denotes a matrix with 1 in all of its entries.
  \item $\uSigma_{3}$ is a diagonal matrix where the $20\%$ of the entries of the diagonal elements are $\sigma^2_j = 0.2p/j$ for $j = 1, \cdots, 0.2p$ and the remaining $\sigma^2_j = 1$ for $j > 0.2p$.
  \item $\uSigma_4$ is an AR(1) covariance matrix with $\sigma_{ij}=\sigma^2\rho^{|i - j|}\bone(|i-j| < 2)$. We chose $\sigma^2 = 1$ and $\rho = 0.4$.
  \item $\uSigma_{5}$ is an AR(1) covariance matrix with $\sigma_{ij}=\sigma^2\rho^{|i - j|}$. We chose $\sigma^2 = 1$ and $\rho = 0.6$.
  \item $\uSigma_6 = \uD^{1/2}\left\{ \uI_K \bigotimes (.2\uI_2 + \uJ_{2}0.8) \right\}\uD^{1/2}$, where $\diag(\uD)=(d_{1}, \cdots, d_{p})\trans$ and $d_{1}, \cdots, d_{p} \sim \text{Uniform}(1, 3)$ and $K = p/2$; $\uI$ is the identity matrix and $\uJ$ a matrix of all ones.
\end{enumerate}
For each case, we also considered two possible alternatives. The mean vectors under the alternative are simulated as follows:
\begin{enumerate} 
  \item[{\bf Alt.1:}] $\umu_g \sim \uN_{p}(\bf{1}, \uI)$, set $p_0$ of its elements to zero and re-scale $\umu_g$ so that $\frac{||\umu_{g}||^{2}}{\sqrt{\trace\left( \uSigma ^{2}\right)}} = 0.1$.
   \item[{\bf Alt.2:}]  $\umu_g \sim \uN_{p}(\bf{1}, \uI)$, set $p_0$ randomly selected elements to zero, and re-scale $\umu_g$ so that $\umu_g \trans\uSigma ^{-1}  \umu_g = 2$.
\end{enumerate}
The two alternatives described above were also previously considered \cite{srivastava2014raptt,zoh2018powerful}.

\subsubsection{Simulation Results}
We first look at the performance of the both tests $\widetilde{\psi}^{PL}$ and $\widetilde{\psi}^{PR}$ in terms of their empirical power for simulation {\bf case 1} under {\bf alternative 1} (See Tables~\ref{tab:table1},~\ref{tab:table2}). 
Overall, both tests tended to have empirical Type 1 error estimates around $5\%$, although in some case the estimated Type 1 error seemed slightly inflated for the case of complex covariance matrices. We note a significant difference between both tests in terms of estimated empirical power.

For the same setting, now looking at {\bf case 2}, the observations made in {\bf case1} still hold (see Tables S.1 ans S.2 from the Supplemental Material), except that we observe a higher estimated power for the test on $\widetilde{\psi}^{PR}$. Recall that in case 2, only the last group had a non-zero mean vector. The test based on the paired groups ($\widetilde{\psi}^{PR}$) performed much better when compared to the test based on the pooled covariance for data simulated under the {\bf alternative 1}. Note that the data were simulated for each group using the same covariance matrix. However, for data simulated under {\bf Alternative 2}, also assuming common group covariance matrices, we see that both the tests based on the pooled covariance ($\widetilde{\psi}^{PL}$) and pairwise groups ($\widetilde{\psi}^{PR}$) performed very similarly (see Tables~\ref{tab:table3}-\ref{tab:table4}). Although the test based on $\widetilde{\psi}^{PL}$ tended to have slightly higher power and estimated Type 1 error near $0.05$ (see Tables~\ref{tab:table3} and ~\ref{tab:table4}).

\begin{table}[ht]
\centering
\caption{ Empirical estimates of the power for the test based on $BF^{PR}$ when data are simulated as in case (1) and based on the Alternative 1. Note here $\Sigma$ refers to the true covariance matrix (G = 3). }
\label{tab:table1}
\resizebox{\columnwidth}{!}{%
\begin{tabular}{|rr|rrrrrr|rrrlll|} \hline
  &&\multicolumn{6}{c|}{ $\uPhi_1$} & \multicolumn{6}{c|}{ $\uPhi_2$} \\ \hline
   & $\Sigma$ &  1 & 0.99 & 0.95 & 0.8 & 0.75 & 0.5 & 1 & 0.99 & 0.95 & 0.8 & 0.75 & 0.5 \\ 
  \hline
    \multirow{6}{*}{\rotatebox[origin=c]{90}{$n=50, p=200$}}
  & 1 &  0.033 & 0.833 & 0.711 & 0.675 & 0.633 & 0.658 &  0.033 & 0.833 & 0.711 & 0.675 & 0.633 & 0.658 \\ 
  & 2 &  0.034 & 0.796 & 0.654 & 0.666 & 0.616 & 0.668 &  0.034 & 0.796 & 0.654 & 0.666 & 0.616 & 0.668 \\ 
    & 3 &  0.038 & 0.780 & 0.643 & 0.610 & 0.586 & 0.589 & 0.038 & 0.780 & 0.643 & 0.610 & 0.586 & 0.589 \\ 
    & 4 &  0.049 & 0.494 & 0.408 & 0.396 & 0.402 & 0.451 &  0.049 & 0.494 & 0.408 & 0.396 & 0.402 & 0.451 \\ 
   & 5 &  0.064 & 0.437 & 0.347 & 0.343 & 0.329 & 0.375 &  0.064 & 0.437 & 0.347 & 0.343 & 0.329 & 0.375 \\ 
    & 6 &  0.073 & 0.283 & 0.235 & 0.224 & 0.217 & 0.237 &  0.073 & 0.283 & 0.235 & 0.224 & 0.217 & 0.237 \\ 
  \cline{3-14} \\
  \cline{3-14}
     \multirow{6}{*}{\rotatebox[origin=c]{90}{$n=70,p=1000$}}&  
     1 & 0.021 & 0.340 & 0.321 & 0.305 & 0.310 & 0.288 &  0.021 & 0.340 & 0.321 & 0.305 & 0.310 & 0.288 \\ 
   & 2 & 0.031 & 0.322 & 0.278 & 0.286 & 0.293 & 0.331 &  0.031 & 0.322 & 0.278 & 0.286 & 0.293 & 0.331 \\ 
    & 3 &  0.062 & 0.292 & 0.262 & 0.237 & 0.237 & 0.233 & 0.062 & 0.292 & 0.262 & 0.237 & 0.237 & 0.233 \\ 
    & 4 &  0.036 & 0.179 & 0.174 & 0.157 & 0.170 & 0.203 & 0.036 & 0.179 & 0.174 & 0.157 & 0.170 & 0.203 \\ 
     & 5 &  0.065 & 0.179 & 0.180 & 0.151 & 0.160 & 0.192 &  0.065 & 0.179 & 0.180 & 0.151 & 0.160 & 0.192 \\ 
   & 6 & 0.047 & 0.143 & 0.126 & 0.104 & 0.119 & 0.134 &  0.047 & 0.143 & 0.126 & 0.104 & 0.119 & 0.134 \\ 
   \hline
\end{tabular}
}
\end{table}

\begin{table}[ht]
\centering
\caption{ Empirical estimates of the power for the test based on $\widetilde{\psi}^{PL}$ when data are simulated as in case (1) and based on the Alternative 1. Note here $\Sigma$ refers to the true covariance matrix (G = 3). }
\label{tab:table2}
\resizebox{\columnwidth}{!}{%
\begin{tabular}{|rr|rrrrrr|rrrlll|} \hline
 & &\multicolumn{6}{c|}{ $\uPhi_1$} & \multicolumn{6}{c|}{ $\uPhi_2$} \\ \hline
   & $\Sigma$ &  1 & 0.99 & 0.95 & 0.8 & 0.75 & 0.5 & 1 & 0.99 & 0.95 & 0.8 & 0.75 & 0.5 \\ 
  \hline
    \multirow{6}{*}{\rotatebox[origin=c]{90}{$n=50, p=200$}} 
 & 1 &   0.034 & 0.050 & 0.042 & 0.033 & 0.051 & 0.045  & 0.034 & 0.050 & 0.042 & 0.033 & 0.051 & 0.045 \\ 
  & 2 &   0.031 & 0.059 & 0.049 & 0.051 & 0.058 & 0.043  & 0.031 & 0.059 & 0.049 & 0.051 & 0.058 & 0.043 \\ 
 & 3 &   0.038 & 0.039 & 0.037 & 0.035 & 0.042 & 0.044  & 0.038 & 0.039 & 0.037 & 0.035 & 0.042 & 0.044 \\ 
   & 4 &   0.049 & 0.064 & 0.053 & 0.067 & 0.059 & 0.055  & 0.049 & 0.064 & 0.053 & 0.067 & 0.059 & 0.055 \\ 
  & 5 &   0.063 & 0.076 & 0.068 & 0.074 & 0.076 & 0.067  & 0.063 & 0.076 & 0.068 & 0.074 & 0.076 & 0.067 \\ 
  & 6 &   0.056 & 0.066 & 0.065 & 0.064 & 0.072 & 0.065  & 0.056 & 0.066 & 0.065 & 0.064 & 0.072 & 0.065 \\ 
   \cline{3-14} \\
  \cline{3-14}
    \multirow{6}{*}{\rotatebox[origin=c]{90}{$n=70,p=1000$}} 
  & 1 &   0.024 & 0.025 & 0.027 & 0.026 & 0.033 & 0.026  & 0.024 & 0.025 & 0.027 & 0.026 & 0.033 & 0.026 \\ 
  & 2 &   0.051 & 0.040 & 0.043 & 0.036 & 0.044 & 0.039  & 0.051 & 0.040 & 0.043 & 0.036 & 0.044 & 0.039 \\ 
  & 3 &   0.065 & 0.071 & 0.068 & 0.050 & 0.056 & 0.061  & 0.065 & 0.071 & 0.068 & 0.050 & 0.056 & 0.061 \\ 
   & 4 &   0.043 & 0.031 & 0.043 & 0.030 & 0.042 & 0.043  & 0.043 & 0.031 & 0.043 & 0.030 & 0.042 & 0.043 \\ 
   & 5 &   0.065 & 0.069 & 0.063 & 0.066 & 0.070 & 0.071  & 0.065 & 0.069 & 0.063 & 0.066 & 0.070 & 0.071 \\ 
  & 6 &   0.049 & 0.036 & 0.058 & 0.053 & 0.042 & 0.066  & 0.049 & 0.036 & 0.058 & 0.053 & 0.042 & 0.066 \\ 
   \hline
\end{tabular}
}
\end{table}

\begin{table}[ht]
\centering
\caption{ Empirical estimates of the power for the test based on $\widetilde{\psi}^{PR}$ when data are simulated as in case (1) and based on the Alternative 2. Note here $\Sigma$ refers to the true covariance matrix (G = 3). }
\label{tab:table3}
\resizebox{\columnwidth}{!}{%
\begin{tabular}{|rr|rrrrrr|rrrlll|} \hline
 & &\multicolumn{6}{c|}{ $\uPhi_1$} & \multicolumn{6}{c|}{ $\uPhi_2$} \\ \hline
   & $\Sigma$ &  1 & 0.99 & 0.95 & 0.8 & 0.75 & 0.5 & 1 & 0.99 & 0.95 & 0.8 & 0.75 & 0.5 \\ 
  \hline
    \multirow{6}{*}{\rotatebox[origin=c]{90}{$n=50, p=200$}} 
 & 1 &  0.033 & 0.570 & 0.450 & 0.429 & 0.408 & 0.408 &  0.033 & 0.570 & 0.450 & 0.429 & 0.408 & 0.408 \\ 
  & 2 & 0.034 & 0.790 & 0.637 & 0.598 & 0.537 & 0.508 & 0.034 & 0.790 & 0.637 & 0.598 & 0.537 & 0.508 \\
  & 3 &  0.038 & 0.989 & 0.997 & 1 & 0.996 & 0.998 &  0.038 & 0.989 & 0.997 & 1 & 0.996 & 0.998 \\ 
 & 4 &  0.049 & 0.716 & 0.603 & 0.563 & 0.518 & 0.518 &  0.049 & 0.716 & 0.603 & 0.563 & 0.518 & 0.518 \\ 
 & 5 & 0.064 & 0.944 & 0.843 & 0.802 & 0.755 & 0.706 &  0.064 & 0.944 & 0.843 & 0.802 & 0.755 & 0.706 \\ 
  & 6 & 0.073 & 0.869 & 0.770 & 0.717 & 0.694 & 0.680 &  0.073 & 0.869 & 0.770 & 0.717 & 0.694 & 0.680 \\ 
    \cline{3-14} \\
  \cline{3-14}
    \multirow{6}{*}{\rotatebox[origin=c]{90}{$n=70,p=1000$}}& 
 1 &  0.021 & 0.704 & 0.655 & 0.628 & 0.635 & 0.606 &  0.021 & 0.704 & 0.655 & 0.628 & 0.635 & 0.606 \\ 
  & 2 &  0.031 & 0.864 & 0.830 & 0.789 & 0.759 & 0.716 &  0.031 & 0.864 & 0.830 & 0.789 & 0.759 & 0.716 \\ 
  & 3 &  0.062 & 1 & 1 & 1 & 1 & 1 &  0.062 & 1 & 1 & 1 & 1 & 1 \\ 
 & 4 &  0.036 & 0.818 & 0.759 & 0.739 & 0.730 & 0.696 &  0.036 & 0.818 & 0.759 & 0.739 & 0.730 & 0.696 \\ 
 & 5 &  0.065 & 0.943 & 0.905 & 0.872 & 0.862 & 0.827 &  0.065 & 0.943 & 0.905 & 0.872 & 0.862 & 0.827 \\ 
 & 6 &  0.047 & 0.899 & 0.859 & 0.823 & 0.818 & 0.794 &  0.047 & 0.899 & 0.859 & 0.823 & 0.818 & 0.794 \\ 
   \hline
\end{tabular}
}
\end{table}

\begin{table}[ht]
\centering
\caption{Empirical estimates of the power for the test based on $\widetilde{\psi}^{PL}$ when data are simulated as in case 1 and based on the Alternative 2. Note here $\Sigma$ refers to the true covariance matrix (G = 3). }
\label{tab:table4}
\resizebox{\columnwidth}{!}{%
\begin{tabular}{|rr|rrrrrr|rrrlll|} \hline
 & &\multicolumn{6}{c|}{ $\uPhi_1$} & \multicolumn{6}{c|}{ $\uPhi_2$} \\ \hline
 & $\Sigma$ &  1 & 0.99 & 0.95 & 0.8 & 0.75 & 0.5 & 1 & 0.99 & 0.95 & 0.8 & 0.75 & 0.5 \\ 
  \hline
    \multirow{6}{*}{\rotatebox[origin=c]{90}{$n=50, p=200$}} 
  & 1 &  0.034 & 0.564 & 0.450 & 0.438 & 0.437 & 0.427  & 0.034 & 0.564 & 0.450 & 0.438 & 0.437 & 0.427 \\ 
 & 2 &  0.031 & 0.781 & 0.674 & 0.590 & 0.596 & 0.530  & 0.031 & 0.781 & 0.674 & 0.590 & 0.596 & 0.530 \\ 
 & 3 &  0.038 & 0.986 & 0.996 & 1 & 0.997 & 0.999  & 0.038 & 0.986 & 0.996 & 1 & 0.997 & 0.999 \\ 
 & 4 &  0.049 & 0.715 & 0.612 & 0.559 & 0.577 & 0.532  & 0.049 & 0.715 & 0.612 & 0.559 & 0.577 & 0.532 \\ 
 & 5 &  0.063 & 0.944 & 0.868 & 0.803 & 0.788 & 0.733  & 0.063 & 0.944 & 0.868 & 0.803 & 0.788 & 0.733 \\ 
 & 6 &  0.056 & 0.872 & 0.783 & 0.731 & 0.739 & 0.708  & 0.056 & 0.872 & 0.783 & 0.731 & 0.739 & 0.708 \\ 
   \cline{3-14} \\
  \cline{3-14}
    \multirow{6}{*}{\rotatebox[origin=c]{90}{$n=70,p=1000$}}
  & 1 &  0.024 & 0.738 & 0.662 & 0.664 & 0.650 & 0.659  & 0.024 & 0.738 & 0.662 & 0.664 & 0.650 & 0.659 \\ 
  & 2 &  0.051 & 0.894 & 0.817 & 0.811 & 0.812 & 0.775  & 0.051 & 0.894 & 0.817 & 0.811 & 0.812 & 0.775 \\ 
  & 3 &  0.065 & 1 & 1 & 1 & 1 & 1  & 0.065 & 1 & 1 & 1 & 1 & 1 \\ 
  & 4 &  0.043 & 0.839 & 0.772 & 0.764 & 0.776 & 0.748  & 0.043 & 0.839 & 0.772 & 0.764 & 0.776 & 0.748 \\ 
  & 5 &  0.065 & 0.947 & 0.916 & 0.889 & 0.910 & 0.859  & 0.065 & 0.947 & 0.916 & 0.889 & 0.910 & 0.859 \\ 
  & 6 &  0.049 & 0.910 & 0.864 & 0.833 & 0.854 & 0.839  & 0.049 & 0.910 & 0.864 & 0.833 & 0.854 & 0.839 \\ 
   \hline
\end{tabular}
}
\end{table}

We also look at the performance of both tests for the case of $5$ groups (see Tables S.3-S.6 of the Supplemental Material). The observation made for the case of $G=3$ groups also holds for the case of $G = 5$.

In the second part of the simulation, we simulated data assuming different covariance matrices between the active group (non-zero) mean vector and the non-active group (all zeros) mean vector. Namely, in {\bf case 1} all groups were assumed to have an identity covariance and the last group $G$ was assumed to have one covariance matrix $\uSigma_k$, for each $k = 1, \cdots, 6$ (see Table~\ref{tab:table5}). 
In {\bf case 2}, however, the active group (Group G) had an identity covariance matrix and the non-active groups had the same covariance matrix which was one of the $\uSigma_k$, for each $k = 1, \cdots, 6$. Under these settings, we see that the test statistic based on $\widetilde{\psi}^{PL}$  (pooled covariance) was poorly calibrated when the covariance matrices were (very) different. So we do not discuss its estimated power here. We instead focus on the test based on $\widetilde{\psi}^{PR}$ (Table~\ref{tab:table6}). The test based on the $\widetilde{\psi}^{PR}$ seemed to hold the nominal Type 1 error for the case of small sample size, even though the estimated Type 1 error seemed inflated when the covariance matrices were very different from the identity matrix. This suggests that large difference between covariance matrices across groups can result in a test statistic that is more liberal. 

\begin{table}[ht]
\centering
\caption{ Empirical estimates of the power for the test based on $\widetilde{\psi}^{PR}$ when data are simulated as in case 2 and based on the Alternative 2. Note here $\Sigma$ refers to the true covariance matrix (G = 3). }
\label{tab:table5}
\resizebox{\columnwidth}{!}{%
\begin{tabular}{|rr|rrrrrr|rrrlll|} \hline
 & & \multicolumn{6}{c|}{ $\uPhi_1$} & \multicolumn{6}{c|}{ $\uPhi_2$} \\ \hline
  & $\Sigma$ &  1 & 0.99 & 0.95 & 0.8 & 0.75 & 0.5 & 1 & 0.99 & 0.95 & 0.8 & 0.75 & 0.5 \\ 
  \hline
  \multirow{6}{*}{\rotatebox[origin=c]{90}{$n=50, p=200$}}
  & 1 &  0.030 & 0.551 & 0.435 & 0.410 & 0.392 & 0.392 &  0.030 & 0.551 & 0.435 & 0.410 & 0.392 & 0.392 \\ 
  & 2 &  0.033 & 0.745 & 0.608 & 0.562 & 0.504 & 0.505 &  0.033 & 0.745 & 0.608 & 0.562 & 0.504 & 0.505 \\ 
  & 3 &  0.039 & 0.997 & 1 & 1 & 0.999 & 1 &  0.039 & 0.997 & 1 & 1 & 0.999 & 1 \\ 
  & 4 &  0.041 & 0.671 & 0.544 & 0.538 & 0.476 & 0.485 &  0.041 & 0.671 & 0.544 & 0.538 & 0.476 & 0.485 \\ 
  & 5 &  0.053 & 0.845 & 0.738 & 0.700 & 0.651 & 0.631 &  0.053 & 0.845 & 0.738 & 0.700 & 0.651 & 0.631 \\ 
  & 6 &  0.086 & 0.980 & 0.942 & 0.925 & 0.915 & 0.914 &  0.086 & 0.980 & 0.942 & 0.925 & 0.915 & 0.914 \\
    \cline{3-14} \\
  \cline{3-14}
    \multirow{6}{*}{\rotatebox[origin=c]{90}{$n=70,p=1000$}}
  & 1 &  0.024 & 0.621 & 0.554 & 0.539 & 0.526 & 0.509 &  0.024 & 0.621 & 0.554 & 0.539 & 0.526 & 0.509 \\ 
  & 2 &  0.015 & 0.811 & 0.763 & 0.714 & 0.708 & 0.667 &  0.015 & 0.811 & 0.763 & 0.714 & 0.708 & 0.667 \\ 
  & 3 &  0.059 & 1 & 1 & 1 & 1 & 1 &  0.059 & 1 & 1 & 1 & 1 & 1 \\ 
  & 4 &  0.011 & 0.738 & 0.692 & 0.659 & 0.659 & 0.617 &  0.011 & 0.738 & 0.692 & 0.659 & 0.659 & 0.617 \\ 
  & 5 &  0.034 & 0.916 & 0.864 & 0.838 & 0.845 & 0.810 &  0.034 & 0.916 & 0.864 & 0.838 & 0.845 & 0.810 \\ 
  & 6 &  0.087 & 1 & 0.999 & 0.998 & 0.999 & 0.997 &  0.087 & 1 & 0.999 & 0.998 & 0.999 & 0.997 \\ 
  \hline
\end{tabular}
}
\end{table}

\begin{table}[ht]
\centering
\caption{ Empirical estimates of the power for the test based on $\widetilde{\psi}^{PR}$ when data are simulated as in case 2 and assuming the Alternative 2. Note here $\Sigma$ refers to the true covariance matrix (G = 3). Here the data is simulated assuming different covariance matrices. }
\label{tab:table6}
\resizebox{\columnwidth}{!}{%
\begin{tabular}{|rr|rrrrrr|rrrlll|} \hline
 & &\multicolumn{6}{c|}{ $\uPhi_1$} & \multicolumn{6}{c|}{ $\uPhi_2$} \\ \hline
 & $\Sigma$ &  1 & 0.99 & 0.95 & 0.8 & 0.75 & 0.5 & 1 & 0.99 & 0.95 & 0.8 & 0.75 & 0.5 \\ 
  \hline
  \multirow{6}{*}{\rotatebox[origin=c]{90}{$n=50, p=200$}}
  & 1 &   0.030 & 0.921 & 0.866 & 0.782 & 0.781 & 0.712 &  0.030 & 0.921 & 0.866 & 0.782 & 0.781 & 0.712 \\ 
 & 2 &   0.033 & 0.993 & 0.966 & 0.922 & 0.931 & 0.881 &  0.033 & 0.993 & 0.966 & 0.922 & 0.931 & 0.881 \\ 
  & 3    & 0.039 & 1 & 1 & 1 & 1 & 1 &  0.039 & 1 & 1 & 1 & 1 & 1 \\ 
  & 4   & 0.041 & 0.979 & 0.952 & 0.885 & 0.910 & 0.821 &  0.041 & 0.979 & 0.952 & 0.885 & 0.910 & 0.821 \\ 
  & 5   & 0.053 & 1 & 0.995 & 0.992 & 0.986 & 0.970 &  0.053 & 1 & 0.995 & 0.992 & 0.986 & 0.970 \\ 
 & 6 & 0.086 & 1 & 0.997 & 0.994 & 0.994 & 0.983 &  0.086 & 1 & 0.997 & 0.994 & 0.994 & 0.983 \\ 
      \cline{3-14} \\
  \cline{3-14}
    \multirow{6}{*}{\rotatebox[origin=c]{90}{$n=70,p=1000$}}
     & 1   & 0.024 & 0.989 & 0.981 & 0.956 & 0.960 & 0.906 &  0.024 & 0.989 & 0.981 & 0.956 & 0.960 & 0.906 \\
  & 2   & 0.015 & 0.999 & 0.999 & 0.995 & 0.994 & 0.973 &  0.015 & 0.999 & 0.999 & 0.995 & 0.994 & 0.973 \\ 
  & 3    & 0.059 & 1 & 1 & 1 & 1 & 1 &  0.059 & 1 & 1 & 1 & 1 & 1 \\ 
    & 4   & 0.011 & 0.996 & 0.993 & 0.989 & 0.980 & 0.966 &  0.011 & 0.996 & 0.993 & 0.989 & 0.980 & 0.966 \\ 
  & 5   & 0.034 & 1 & 1 & 1 & 1 & 1 &  0.034 & 1 & 1 & 1 & 1 & 1 \\ 
  & 6   & 0.087 & 1 & 1 & 1 & 1 & 1 &  0.087 & 1 & 1 & 1 & 1 & 1 \\ 
  \hline
\end{tabular}
}
\end{table}

\section{Application} \label{sec:Application}
The data set used in our application originated from a head and neck squamous cell carcinoma (HNSCC) study where the profiles of $5902$ single cells were obtained from 18 patients with oral cavity tumors by single cell RNA-seq \cite{puram2017single}. The data set used for our analysis can be downloaded from the Gene Expression Omnibus (\textbf{\url{https://www.ncbi.nlm.nih.gov/geo/query/acc.cgi?acc=GSE103322}}). Each of the $5902$ cells were identified and labelled. 
We wanted to know whether there is evidence that cell types have different gene expression profiles while accounting for the potential dependency between the genes. We framed this problem as a high-dimensional mean vector test. Under the null hypothesis, all cell types have equal or similar mean gene expression profiles. Before we applied our proposed test, we performed a feature reduction step to root out low expressed genes. Similar to a prior approach \cite{puram2017single}, we chose genes with $E_a(i) > 6$, where $E_a(i)=\log2\text{(average(TPM(i)1...k)+1)}$. This resulted in $p = 2641$ genes selected for analysis. We consider the less abundant (tumor free) cell types (3 types): B-cells (n=138), macrophages (n=98), and mast (n=120). To perform our test, we selected the dimension of the projections space, according to our discussion above, to be $m = 88$ (for the test based on $\widetilde{\psi}^{PR}$) and $m = 161$ for the test based on $\widetilde{\psi}^{PL}$. We selected $\tau^{\star}_{ij}$ and $\gamma^{\star}_{ij}$ according to our discussion in Section~\ref{sec:testmtaugam} and set $\alpha$ to $.05$. Since our test might be sensitive to significant departures from the assumption of common covariance, we performed tests comparing covariance matrices \cite{ahmad2017location,srivastava2010testing}. Interestingly enough, the null hypothesis was rejected ($p-value < 0.001$) with one test \cite{ahmad2017location} but not another ($p-value = 0.1878$) \cite{srivastava2010testing}. 
Finally, applying our test to the data set, we rejected the null hypothesis with an estimated $p-value < 0.001$ based on the test statistic approximated by assuming zero mean vectors for each group and a common identity covariance matrix (\ref{eq:BFensblPL}, \ref{eq:BFensblPR}). The tests based on the Bayes factor assuming a common overall covariance matrix ($\widetilde{\boldpsi}^{PL}$) and a pairwise Bayes factor $(\widetilde{\boldpsi}^{PR})$ both yielded the same result for both projection matrices $\uPhi_1$ and $\uPhi_2$. 
Our testing procedure also provides an automatic way to extract information about all pairwise comparisons since the value of each $f_{ij}$ statistic is retained. A useful summary statistic we can look at is the proportions of the $f_{ij}$ statistic that exceeded the threshold of significance across all random projections (Table~\ref{tab:table7a}). We note that the test assuming common covariance across all groups $\widetilde{\boldpsi}^{PL}$ tended to have a larger proportion of significant tests for all pairwise comparisons when compared to its $\widetilde{\boldpsi}^{PR}$ counterpart. Overall, we conclude that the three cell types have different gene expression profiles, further justifying why they are clustered as different cell types.  

\begin{table}[ht]
\centering
\caption{Proportion of pairwise tests that were declared significant across $1000$ random projections. The results are reported for the both tests $(BF^{PR}_{10}, BF^{PL}_{10})$ .}
\label{tab:table7a}
\begin{tabular}{|rrr|}
  \hline
   & Macrophage & B Cell \\
    Mast & (0.763, 0.943)  & (0.689, 0.819) \\
    Macrophage & - & (0.987, 1.0)  \\
  \hline
\end{tabular}
\end{table}




\section{Conclusion} \label{sec:conclusion}
When the dimension of the feature space exceeds the combined sample size, classical MANOVA test statistics cannot be used to compare multiple group means and some regularization steps are needed. We addressed the problem of multiple group mean vector testing with random projections(RPs). We formulated two tests based on Bayes factors with different assumptions about the covariance matrix. In one test, we assumed one overall covariance matrix which results in a Bayes factor denoted as $BF^{PL}_{}(\widetilde{\psi}^{PL})$. In the other test, we assumed only a pairwise common covariance matrix with the Bayes factor denoted as $BF^{PR}_{}(\widetilde{\psi}^{PR})$. When the assumption of a homogeneous covariance matrix was reasonable, both test statistics performed similarly and very well. However, for moderate departures from the assumption of common covariance, the test based on $BF^{PR}_{}$ seemed robust in the simulation setting we considered. Although, we should note that the test based on $BF^{PR}_{}$ seemed to have an inflated estimated type 1 error in some settings. A natural extension of this work is to address the case of potentially very different covariance matrices. Additionally, our test statistic was derived assuming a normal distribution, so it will be interesting to relax that potentially restricted assumption. Our approach was implemented in the  \textsf{Julia} \cite{bezanson2012julia} statistical software and will be made available for used on the first author's Github page.   

\section*{Acknowledgments}

We thank the Editor, Associate Editor and referees. This research was supported by diversity supplements under award numbers U01-CA057030-29S1 and by Lilly Endowment, Inc., through its support for the Indiana University Pervasive Technology Institute.





\bibliographystyle{myjmva}
\bibliography{Bibliography_lrtnew,Bibliography_lrtnew2,Corrected_Bib2}
\end{document}